\documentclass[10pt,journal]{IEEEtran}
\IEEEoverridecommandlockouts
\usepackage{cite}
\usepackage{enumitem}
\usepackage{amsmath,amssymb,amsfonts}
\usepackage{graphicx}
\usepackage{amsmath}
\usepackage{xcolor}
\usepackage{multirow}
\usepackage{amssymb}
\newcommand\myworries[1]{\textcolor{blue}{#1}}
\usepackage{amsthm}
\newtheorem{remark}{Remark} %\newtheorem* --> for no number of remark.

\newtheorem*{theorem*}{Theorem}

\newtheorem*{lemma*}{Lemma}
\usepackage{tabularx}
\usepackage{url}

\usepackage{array}
\newcolumntype{P}[1]{>{\centering\arraybackslash}p{#1}}

\usepackage{enumitem}
\newlist{legal}{enumerate}{10}
\setlist[legal]{label*=\arabic*.}

\newtheorem*{definition*}{Definition}

\newtheorem*{condition*}{Condition}
\newtheorem*{proof*}{Proof}

\renewcommand\myworries[1]{}
\usepackage{mathtools}
\newsavebox\myboxA
\newsavebox\myboxB
\newlength\mylenA

\newcommand{\subparagraph}{}

\newcommand*\xoverline[2][0.75]{%
	\sbox{\myboxA}{$\m@th#2$}%
	\setbox\myboxB\null% Phantom box
	\ht\myboxB=\ht\myboxA%
	\dp\myboxB=\dp\myboxA%
	\wd\myboxB=#1\wd\myboxA% Scale phantom
	\sbox\myboxB{$\m@th\overline{\copy\myboxB}$}%  Overlined phantom
	\setlength\mylenA{\the\wd\myboxA}%   calc width diff
	\addtolength\mylenA{-\the\wd\myboxB}%
	\ifdim\wd\myboxB<\wd\myboxA%
	\rlap{\hskip 0.5\mylenA\usebox\myboxB}{\usebox\myboxA}%
	\else
	\hskip -0.5\mylenA\rlap{\usebox\myboxA}{\hskip 0.5\mylenA\usebox\myboxB}%
	\fi}
\makeatother

\usepackage{scalerel}[2014/03/10]
\usepackage{stackengine}

\newlength\myindent
\usepackage{epsfig}
\usepackage{tabularx}
\usepackage{supertabular,booktabs}
\usepackage{capt-of}

\usepackage{longtable}
\usepackage{siunitx}
\usepackage{booktabs}

\usepackage{subcaption}

\usepackage{algorithm}
\usepackage{algpseudocode}

\newcommand{\bd}{\mathbf}

\algrenewcommand\algorithmicrequire{\textbf{Precondition:}}
\algrenewcommand\algorithmicensure{\textbf{Postcondition:}}

\usepackage{etoolbox}
\usepackage{textcomp}
\usepackage[normalem]{ulem}
\usepackage{mathtools}
\usepackage{float}
\usepackage{tikz}
\usetikzlibrary{shapes.geometric, arrows}
\tikzstyle{startstop} = [rectangle, rounded corners, minimum width=3cm, minimum height=1cm,text centered, draw=black, fill=red!30]
\tikzstyle{io} = [trapezium, trapezium left angle=70, trapezium right angle=110, minimum width=3cm, minimum height=1cm, text centered, text width=3cm, draw=black, fill=blue!30]
\tikzstyle{process} = [rectangle, minimum width=3cm, minimum height=1cm, text centered, draw=black, fill=orange!30]
\tikzstyle{decision} = [diamond, minimum width=2.5cm, minimum height=2.5cm, text centered, draw=black, fill=green!30]
\tikzstyle{arrow} = [thick,->,>=stealth]
\tikzstyle{process} = [rectangle, minimum width=3cm, minimum height=1cm, text centered, text width=3cm, draw=black, fill=orange!30]
\def\BibTeX{{\rm B\kern-.05em{\sc i\kern-.025em b}\kern-.08em
		T\kern-.1667em\lower.7ex\hbox{E}\kern-.125emX}}
		
\usepackage[compact]{titlesec}
\titlespacing{\section}{0pt}{*0}{*0}
\titlespacing{\subsection}{0pt}{*0}{*0}
\titlespacing{\subsubsection}{0pt}{*0}{*0}
	
\begin{document}
	\bstctlcite{IEEEexample:BSTcontrol}
    
	% Distributed WAM based voltage stability monitoring using powerflow circles
	%Analytically Exact Distributed Voltage Stability Monitoring Scheme Using Power Flow Circles
	\title{PMU-based Distributed Non-iterative Algorithm for Real-time Voltage Stability Monitoring\\
		 
		\thanks{K. P. Guddanti and Y. Weng are with the School of Electrical, Computer and Energy Engineering at Arizona State University, emails:\{kguddant,yang.weng\}@asu.edu; A. R. R. Matavalam is with Department of Electrical and Computer Engineering at Iowa State University, email: amar@iastate.edu.} \thanks{\copyright 2020 IEEE.  Personal use of this material is permitted.  Permission from IEEE must be obtained for all other uses, in any current or future media, including reprinting/republishing this material for advertising or promotional purposes, creating new collective works, for resale or redistribution to servers or lists, or reuse of any copyrighted component of this work in other works.} \thanks{\textbf{\underline{Full citation}}: Guddanti, Kishan Prudhvi, Amarsagar Reddy Ramapuram Matavalam, and Yang Weng. "PMU-Based Distributed Non-Iterative Algorithm for Real-Time Voltage Stability Monitoring." IEEE Transactions on Smart Grid 11, no. 6 (2020): 5203-5215. DOI: 10.1109/TSG.2020.3007063} \thanks{\textbf{\textit{This is the first method in the literature that is NOT Thevenin and NOT sensitivity type of approach in the literature while retains the advantages of both Thevenin and Sensitivity based methods together.}}}
	}

\makeatletter	
\patchcmd{\@maketitle}
  {\addvspace{0.5\baselineskip}\egroup}
  {\addvspace{-1\baselineskip}\egroup}
  {}
  {}
%	Kishan Prudhvi Guddanti, Yang Weng, Amarsagar Reddy Ramapuram Matavalam,Student Member

	\author{
		\IEEEauthorblockN{Kishan Prudhvi Guddanti},
		\IEEEauthorblockA{\textit{Student Member}},
		\textit{IEEE},
		\and
		\IEEEauthorblockN{Amarsagar Reddy Ramapuram Matavalam},
		\IEEEauthorblockA{\textit{Member}},
		\textit{IEEE},
		\and
		\IEEEauthorblockN{Yang Weng},
		\IEEEauthorblockA{\textit{Member}},
		\textit{IEEE}
	   % \and 
	   % \\ [-10.0em] 
	}
	%\DeclarePairedDelimiter\abs{\lvert}{\rvert}
	%\makeatletter
	%\let\oldabs\abs
	%\def\abs{\@ifstar{\oldabs}{\oldabs*}}
	
	%\IEEEaftertitletext{\vspace{-1\baselineskip}}
	
	\maketitle
%--------------------------------------------------------------------------------------------------------------------------	
	\begin{abstract}
	The Phasor measurement unit (PMU) measurements are mandatory to monitor the power system's voltage stability margin in an online manner. Monitoring is key to the secure operation of the grid. Traditionally, online monitoring of voltage stability using synchrophasors required a centralized communication architecture, which leads to the high investment cost and cyber-security concerns. The increasing importance of cyber-security and low investment costs have recently led to the development of distributed algorithms for online monitoring of the grid that are inherently less prone to malicious attacks. In this work, we proposed a novel distributed non-iterative voltage stability index (VSI) by recasting the power flow equations as circles. The processors embedded at each bus in the smart grid with the help of PMUs and communication of voltage phasors between neighboring buses perform simultaneous online computations of VSI. The distributed nature of the index enables the real-time identification of the critical bus of the system with minimal communication infrastructure. The effectiveness of the proposed distributed index is demonstrated on IEEE test systems and contrasted with existing methods to show the benefits of the proposed method in speed, interpretability, identification of outage location, and low sensitivity to noisy measurements.	
	\end{abstract}
	
	%\vspace{-4mm}
	\section{Introduction}
%	\input{Intro.tex}
	%--------------------------------------------------------------------------------------------------------------------------

The advent of phasor measurement units (PMUs) and their wide-spread acceptance has made it possible to obtain real-time and time-synchronized information of voltage and current phasors for use in diversified applications \cite{Phadke93,cui2019enhance}. Consequently, the proliferation of PMUs transformed the power grid into a tightly integrated cyber-physical system, where local computation is viable for distributed monitoring and coordinated control actions in real-time for wide-area monitoring, protection, and control (WAMPAC).

Among WAMPAC applications, online monitoring of long-term voltage instability (LTVI) is an area of interest for industry \cite{Novosel08}. Specifically, LTVI is a quasi-static bifurcation (nose point of the PV curve) \cite{Dobson92}, caused by the inability of the generation and transmission system to provide sufficient power to loads, e.g., due to increased demand, generation outage, or generators on VAR limits \cite{Cutsem98,Ajjarapu07}. If left unattended, LTVI results in a system-wide voltage collapse and blackout \cite{bollen03,taylor00}. Using PMU measurements, we can compute the margin (a measure of distance) from the current operating condition to LTVI. This measure of distance is quantified using metrics known as \textit{voltage collapse proximity indices} (VCPIs). The accuracy of VCPIs depends on the theoretical basis used to derive them, e.g., Thevenin circuit, sensitivity methods \cite{Chebbo92,Haque95}. These methods compute the VCPIs by either decentralized(local) or centralized communication architectures.

Conventional centralized VCPIs typically use Jacobian-based sensitivity indices. These methods use Jacobian to identify LTVI since the Jacobian matrix becomes singular at LTVI  \cite{Amar18,Weng}. However, the calculation of the Jacobian requires voltage phasor information at every bus in the power system, which is expensive \cite{Wang11,Liu14,Glavik091,Glavik092}. In addition to this, the interpretability of sensitivity based VCPIs is harder than decentralized(local) VCPIs. It is hard because, when the operating point is close to the limit, the corresponding centralized VCPI value suddenly increases, becoming unbounded. The unbounded nature and non-linear behavior make the centralized VCPI hard to interpret as a fair distance to the critical loading. References \cite{Kessel86,Wang11,Liu14} address the non-interpretability partially in a \textit{centralized manner} using measurements from all the nodes in the system. Specifically, the original $n$-bus system is simplified into a $2$-bus multi-port coupled network equivalents. However, there is no theoretical proof that the LTVI of the full-system can always be identified using these simplifications.

To overcome the drawbacks of centralized methods, methods (indices) that utilize only local PMU measurements at a substation are developed to monitor the LTVI locally, and these indices are referred to as decentralized VCPIs \cite{Vu99}. Such VCPIs employ Thevenin-based methods by utilizing the quasi-steady-state behavior of the power system to calculate a $2$-bus Thevenin equivalent at a bus \cite{Vu99}. In such an equivalent model, one node represents the bus of interest while the other accounts for the rest of the system. The equivalent circuit estimation only requires local measurements (i.e., requires only a single PMU) at a bus of interest over a time period, making it a decentralized architecture. Once the equivalent circuit is estimated, an index value is computed at the bus of interest using maximum power transfer theorem \cite{Vu99}. Therefore, these Thevenin-based methods have a nice property of decentralized computation of the index values \cite{Gubina95,Vu99,Milosevic03,Verbic04}. It is also shown that the local index can theoretically (in the presence of no measurement noise) always identify the LTVI of the full system by correlating it with the Jacobian \cite{Amar18}. Interpretability is an additional advantage of decentralized VCPIs, as the values are usually between $0$ and $1$. Unfortunately, the decentralized method suffers from errors in practice, especially due to the presence of measurement noise, e.g., the local VCPIs can be inaccurate with large errors \cite{Zima05}.

\begin{figure}[H]
    \centering
    \centerline{\includegraphics[width=\linewidth]{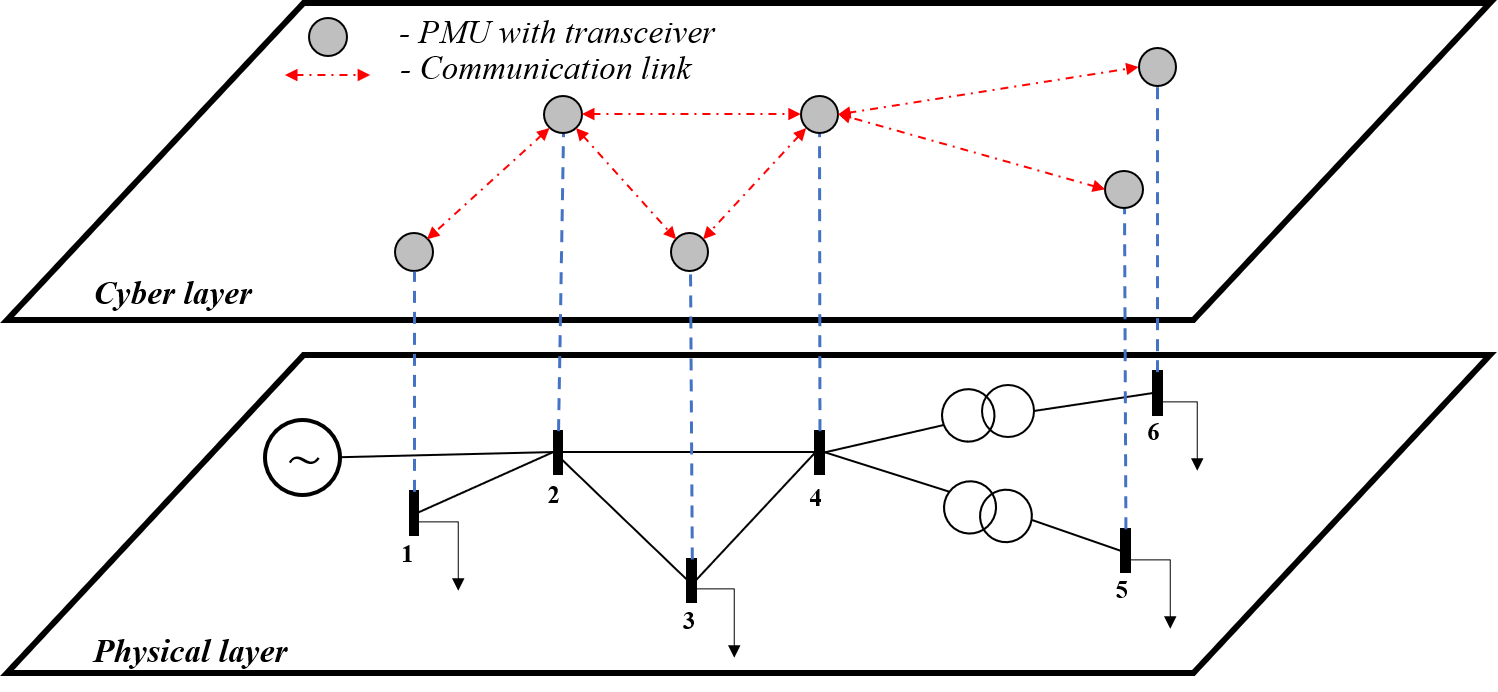}}
    \caption{Cyber-physical form of a futuristic power system with communication links. The PMUs cover all nodes in this cyber layer and only communicate between neighbours. This is when VCPI of every node needs to be computed. In a scenario where the operator would like monitor only a few critical nodes that are important, only a few PMUs are needed as shown in Fig.~\ref{fig:cyber_physical_system_2}.}
    \label{fig:cyber_physical_system}
\end{figure}

As the centralized and decentralized methods have trade-offs between accuracy, cost, and interpretability, we are interested in an intermediate scheme to include these advantages without minimal trade-offs. Specifically, the centralized method requires measurements from the entire system, while the decentralized method only requires the PMU measurement on the bus of interest. In this paper, we propose to use an intermediate approach that requires PMU measurements from the neighboring buses of the interested bus to calculate the proposed VCPI. This set up is known as a distributed monitoring scheme, as shown in Fig.~\ref{fig:cyber_physical_system}. In this scheme, PMUs communicate their measurements at a bus in the physical layer with its adjacent buses via the cyber layer. The processor embedded at each bus only uses its adjacent buses' voltage phasor measurements to compute the proposed VCPI referred to as voltage stability index ({VSI}).  

The structure of the power system in Fig.  \ref{fig:cyber_physical_system} {is for a future grid communication architecture. Some of these communication advantages are robustness to cyber-attacks on the communication nodes, and improved reliability of communications due to shorter communication links \cite{baran1964distributed,wang14decentralized}. Hence it is highly desired to have a distributed communication architecture due to its various advantages in contrast to star/centralized architecture.} Hence it has been of interest in the community recently for distributed implementation of various applications. For example, WAMPAC \cite{Domiguez11,Porco16,weng2016distributed}, load shedding \cite{Xu11, Xu18}, optimal power flow \cite{magnusson15}, economic dispatch \cite{Dorfler16}, transfer capability assessment \cite{Liu15}, and control techniques in microgrids \cite{Porco15,Wang17}. 
Even for VCPI, \cite{Porco16} uses such a distributed communication setup, where the Jacobian is used to calculate a sensitivity index iteratively. Unfortunately, the advantage of the trade-off between centralized method and the decentralized method is not utilized fully, making \cite{Porco16} suffer from $1$) necessity for PMUs to cover all buses, $2$) increasing computational time as the power system operating point gets closer to critical loading, $3$) non-interpretable nature of the index, $4$) theoretical flaw of assuming eigenvalues of the Jacobian to be non-negative (invalid in the IEEE $300$-bus network).

%--------------------------------------------------------------------------------------------------------------------------
\textbf{Contributions}: In this paper, we propose a novel distributed index that overcomes the drawbacks of \cite{Porco16} mentioned above. Instead of exploiting the properties of the Jacobian, we analyzed the power flow equations themselves with a new perspective as circles in a distributed framework. This leads to a new voltage stability index (VSI) that {$1$) does not need PMUs at all buses $2$) is non-iterative leading to fast computation irrespective of system loading, $3$) is interpretable (normalized between $1$ and $0$), where $1$ indicates no-load and $0$ indicates maximum load, $4$) is exact (assumption free) unlike other indices in literature \cite{Vu99,Kessel86,Wang11,Liu14}. This is the first time that a non-iterative, approximation free as well as distributed communication based method has been proposed for monitoring the LTVI.}

%--------------------------------------------------------------------------------------------------------------------------

The paper is organized as follows: Section~\ref{sec:PF_eqs_review} explains the power flow equations as circles in a distributed framework. Section~\ref{sec:Distributed_VSI} quantifies the distance between the power flow circles as a measure of margin to LTVI. Section~\ref{sec:Simulations} shows the efficacy of the proposed scheme and VSI via simulations on different IEEE systems. Section~\ref{sec:conclusion} concludes the paper.
%--------------------------------------------------------------------------------------------------------------------------

%--------------------------------------------------------------------------------------------------------------------------
	%--
	\section{Power Flow Equations as Intersection of Circles}
	\label{sec:PF_eqs_review}
	Let $p_d$ and $q_d$ be the active and reactive power injections at bus $d$. Let  $\mathit{g}_{k,d} + j\cdot \mathit{b}_{k,d}$ be the $(k,d)^{th}$ element of the admittance matrix $Y$. $\mathit{v}_{d,r}$ and $\mathit{v}_{d,i}$ be the real and imaginary part of the voltage phasor at bus $d$, respectively. We represent bus $d$'s neighboring bus set as $\mathcal{N}(\mathit{d})$. The power flow equations used in this paper are in rectangular coordinate form. They are given by
\begin{subequations}\label{eqn:pf_eqs}
\begin{equation}
\mathit{p}_d
= t_{d,1} \cdot \mathit{v}^2_{d,r} + t_{d,2} \cdot \mathit{v}_{d,r} + t_{d,1} \cdot \mathit{v}^2_{d,i} + t_{d,3} \cdot \mathit{v}_{d,i},    
\label{math_p}
\end{equation}
\begin{equation}
\mathit{q}_d
= t_{d,4} \cdot \mathit{v}^2_{d,r } - t_{d,3} \cdot \mathit{v}_{d,r} + t_{d,4} \cdot \mathit{v}^2_{d,i} + t_{d,2} \cdot \mathit{v}_{d,i}.    
\label{math_q}
\end{equation}
\end{subequations}
The parameters $t_{d,1},t_{d,2},t_{d,3}$ and $t_{d,4}$ are given by
\begin{subequations}\label{eqn:parameters}
\begin{equation}
t_{d,1} 
= - \sum_{\mathit{k}\in\mathcal{N}(\mathit{d})} \mathit{g}_{k,d} ,
\ t_{d,2}
= \sum_{\mathit{k}\in\mathcal{N}(\mathit{d})} (\mathit{v}_{k,r}\mathit{g}_{k,d} - \mathit{v}_{k,i}\mathit{b}_{k,d}) ,
\label{ref1}
\end{equation}
\begin{equation}
t_{d,3}
= \sum_{\mathit{k}\in\mathcal{N}(\mathit{d})} (\mathit{v}_{k,r}\mathit{b}_{k,d} + \mathit{v}_{k,i}\mathit{g}_{k,d}) ,
\ t_{d,4} 
=  \sum_{\mathit{k}\in\mathcal{N}(\mathit{d})} \mathit{b}_{k,d}.
\label{ref2}      
\end{equation}
\end{subequations}
\subsection{Distributed Nature of Power Flow Equations as Circles}
\label{subsec:dist_nature_of_pf_circles}
\begin{figure}[H]
    \centering
    \centerline{\includegraphics[width=\linewidth]{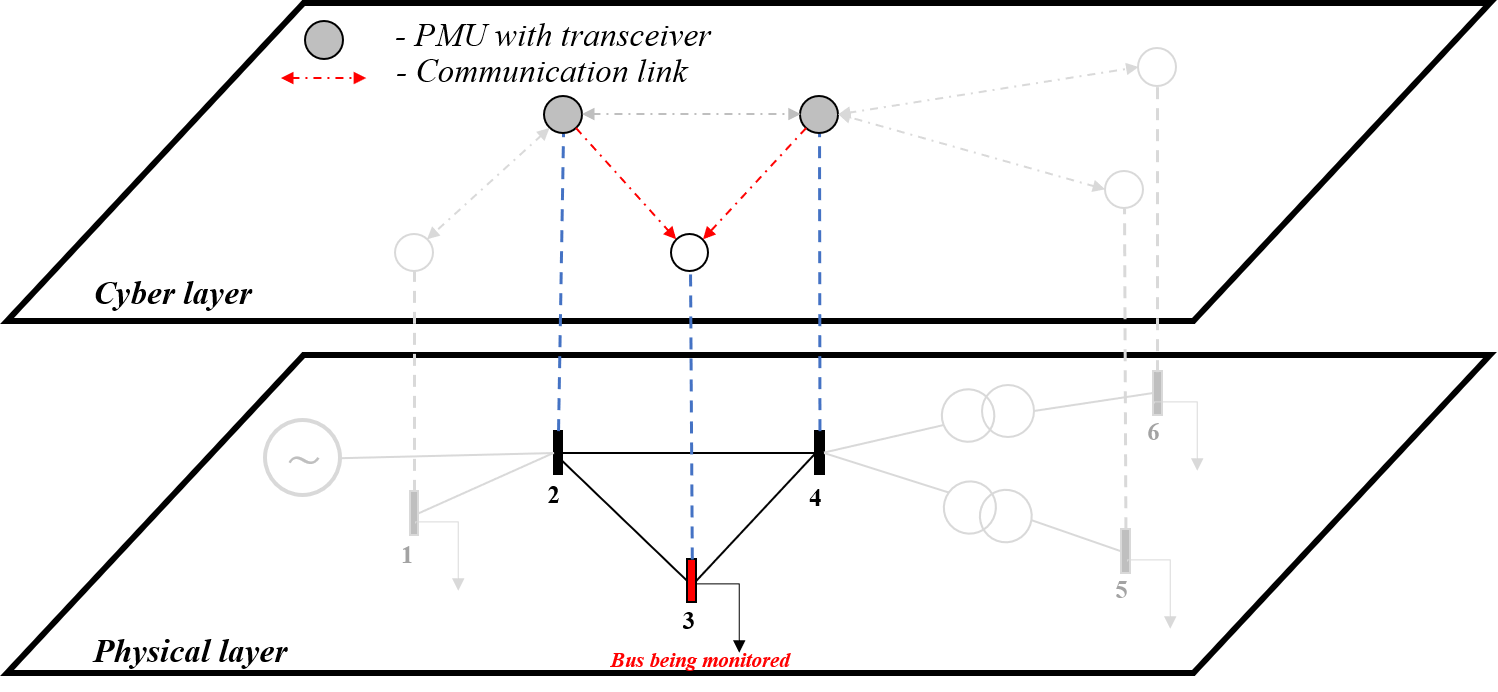}}
    \caption{Cyber-physical form of the power system to monitor {the VCPI of} bus $3$ in the network using proposed method. To monitor the margin at bus $3$ in an online fashion, we only need the PMU measurements from its adjacent buses i.e., buses $2$, $4$ {and branch admittance of lines joining bus $3$ to $2$ and $4$. We do not need the admittance matrix of entire system.}}
    \label{fig:cyber_physical_system_2}
\end{figure}

The advantage of using power flow equations \eqref{eqn:pf_eqs} comes from their visualization as circles. Specifically, for fixed constants $t_{d,1}, t_{d,2}, t_{d,3}, t_{d,4}$, equations \eqref{math_p} and \eqref{math_q} represents two circles at bus $d$ in $v_{d,r}$ and $v_{d,i}$ space \cite{Weng}. The real power equation \eqref{math_p} at bus $d$  can be represented as a circle with center {${o}_{p}$} and radius $\mathit{r}_{p}$. The reactive power equation \eqref{math_q} at bus $d$  can be represented as a circle with center {${o}_{q}$} and radius $\mathit{r}_{q}$. These centers and radii are given by
\begin{subequations}\label{eqn:crpq}
\begin{align}
&{{o_p}}
= \left(\frac{-t_{d,2}}{2t_{d,1}}, \frac{-t_{d,3}}{2t_{d,1}}\right), \;
{{o}_{q}}
= \left(\frac{t_{d,3}}{2t_{d,4}}, \frac{-t_{d,2}}{2t_{d,4}}\right), \label{eq:cpq}\\
&\mathit{r}_{p}
= \sqrt{\frac{\mathit{p}_{d}}{t_{d,1}} +  \frac{\left(t_{d,2}\right)^2 + \left(t_{d,3}\right)^2}{4t^2_{d,1}}}, \label{eq:rp}\\
&\mathit{r}_{q}
= \sqrt{\frac{\mathit{q}_d}{t_{d,4}} +  \frac{\left(t_{d,3}\right)^2 + \left(t_{d,2}\right)^2}{4t^2_{d,4}}}.\label{eq:rq}
\end{align}
\end{subequations}
Using the centers and radii \eqref{eqn:crpq}; the intersection points of real and reactive power circles at bus $d$ provides the voltage solution $v_d$ at bus $d$ where $v_d = v_{d,r}+j\cdot v_{d,i}$. It is important to note that the expressions for the centers and radii of the power flow circles at bus $d$ contain $t_{d,2}$ and $t_{d,3}$ which contain the neighboring bus voltages $v_{k,r}+j \cdot v_{k,i}$ where $\mathit{k}\in\mathcal{N}(\mathit{d})$. To plot the power flow circles at bus $d$, the centers and radii in \eqref{eqn:crpq} are calculated locally by the embedded processor at bus $d$ using the voltage phasor measurements from the PMUs located at the immediate neighboring buses to bus $d$. For example, to plot the power flow circles at bus $3$ in Fig.~\ref{fig:cyber_physical_system_2}, we only need the voltage phasors from the PMUs at buses $2$ and $4$; branch admittance of lines joining bus $3$ to $2$ and $4$. Fortunately, in the proposed distributed scheme, we do not need the admittance matrix of the entire system. 

The discussion above is for PQ buses with constant real and reactive power constraints. In {the} case of PV buses, we get constant real power and voltage constraints{,} which also lead to two circles, namely the real power circle \eqref{math_p} and voltage circle centered at {the} origin with radius $\left|V_{\textrm{specified}}\right|$. We will next discuss how the power flow circles are impacted due to load increase. 
\subsection{Impact of Load Increase on Power Flow Circles}
\label{subsec:impact_load_incr_on_pf_circles}
The long-term voltage instability is caused when the generation and transmission system cannot supply power demanded by the load. This situation corresponds to the lack of a solution for the power flow equations. This can be identified by looking at the intersection of power flow circles constructed using PMU measurements at bus $d$. Geometrically, two circles can have two, one or no points of intersections. Hence, the voltage solution at bus $d$ represented as an intersection of power flow circles can have (a) two common points to indicate multiple feasible voltage solutions, or (b) one common point to indicate a single feasible voltage solution corresponding to the nose point (LTVI) of PV curve, or (c) no common points to represent the in-feasibility of operating conditions. 

Before we provide an example, here we present a high-level data-flow of any voltage stability monitoring method and explain the overall procedure for the calculation of a VCPI. The sequential steps for any online stability monitoring method are: (a) Collection of real-time PMU measurements from the grid, (b) Estimation of power system states from the PMU measurements, and (c) Utilizing the estimated states to calculate an index that provides a metric of the system stability (VCPI). At a given time step or ``snapshot", we are monitoring the voltages in the power grid directly from the PMUs. When there is a load change in the power grid changes in the next snapshot, the PMU voltage measurements in the next snapshot are automatically updated since PMU device voltages are functions of loading on the power grid. Using these real-time PMU measurements directly the proposed index is calculated. Alternatively, the input measurements to calculate the proposed index can also be obtained from the output of a preprocessor such as a state estimator.

\begin{figure}
    \centering
    \centerline{\includegraphics[scale=0.3]{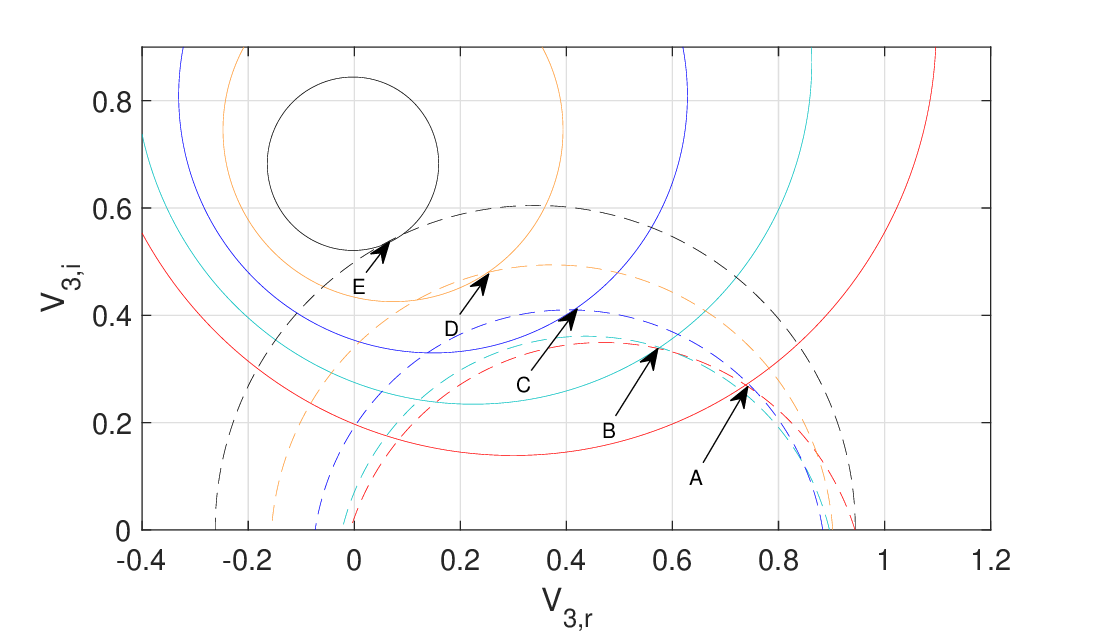}} %width=\linewidth
    \caption{Power flow circles at bus $3$ with the voltage solutions for increasing loads.}
    \label{fig:Zoom_circle_plot}
\end{figure}

For illustration purpose, we use a completely connected $3$-bus system with the same branch admittance value of $1-j\cdot0.5$ p.u. To showcase the behavior of power flow circles due to the impact of load change, we will increase the loading on the power grid until there is a blackout and observe the power flow circles' behavior. We also assume the same apparent load at bus $2$ and $3$, e.g., $S_2 = S_3$ for simplicity. The loads at bus $2$ and $3$ are increased in $5$ different time instants with the following values (in p.u.) $[-0.01+j\cdot 0.33,-0.04+j\cdot 0.40,-0.13+j\cdot 0.44,-0.28+j\cdot 0.45,-0.49+j\cdot 0.43]$ until the power flow circles at bus $3$ had only one common point. To calculate the power flow circles' radius and center corresponding to the $5$ different time instants in real-time, we use the distributed communication architecture described in Fig.~\ref{fig:cyber_physical_system_2} to directly collect the real-time PMU voltage measurements for the $5$ different time instants. Using these neighboring bus PMU voltages and \eqref{eqn:crpq}, the circles can be drawn in real-time. The corresponding power flow circles' intersection is plotted as shown in %Fig.~\ref{fig:circle_plot} and 
Fig.~\ref{fig:Zoom_circle_plot}, where the solid and dashed circles indicate the reactive and real power circles respectively.

In Fig.~\ref{fig:Zoom_circle_plot}, the common points ``A", ``B", ``C", ``D", and ``E" indicate the higher magnitude voltage solutions corresponding to an apparent power load of $[-0.01+j\cdot 0.33,-0.04+j\cdot 0.40,-0.13+j\cdot 0.44,-0.28+j\cdot 0.45,-0.49+j\cdot 0.43]$ p.u. respectively at {the} bus $3$. It can be observed that the reactive power circle steadily becomes smaller{,} and the real power circle steadily becomes larger as the system load increases. This behavior of power flow circles shows that the circles move further away from each other until they have only one common point (nose point) as the load increases to the critical value. This occurs in Fig.~\ref{fig:Zoom_circle_plot} at the intersection of black power flow circles. It corresponds to the critical loading in the system and has only one intersection point ``E". This point represents the nose point of {the} PV curve. It is to be noted that all the power flow circles at every bus will not touch each other externally at a given snapshot, but only the power flow circles at the bus that is critically loaded will touch externally indicating the LTVI.
Thus, the distance between the power flow circles at the critical bus decreases to zero as the load increases to the critical value. Hence, this distance can be used as an indicator {of} LTVI. Quantifying the distance between the circles for deriving a VSI is described in the next section. 

%-------------------------------------------------------------------------------------------------------------------------
	%--
	\section{Characterizing the Distance Between Power Flow Circles for VSI}
	\label{sec:Distributed_VSI}
	%###################################################################################################
The distance between the power flow circles constructed using the PMU measurements indicates the margin to reach LTVI. There are several ways to quantify the distance between two circles. One such indicator may consider the distance between the circles using their centers and radii. This distance is positive, zero and negative when the circles intersect, touch externally and do not intersect with each other{,} respectively. Another indicator is the Euclidean distance between the two common points when the circles intersect each other. When the circles intersect, touch externally{,} and do not intersect with each other, the Euclidean distance between the two voltage solutions is positive, zero{,} and does not exist{,} respectively. {We observed that the proposed index (VSI) derived below in contrast to the above mentioned possible formulations of distributed indices has a relatively better behavior with changing loading.}

Different than the {indicators mentioned above}, the proposed index is derived by using the concept of {the} family of circles instead of directly using \eqref{math_p} and \eqref{math_q}. In this section first, we introduce the concept of determinant to infer the intersection of any two circles. Second, using the determinant, we derive the expression for the proposed distributed voltage stability index.

\subsection{Identifying the Intersection of Power Flow Circles Using Determinant}
\label{subsec:quant_dist_btwn_pf_circles}
Given real and reactive power circles, two important aspects have to be inferred to derive the proposed indicator. They are 1) the existence of {a} feasible solution{,} i.e., does the power flow circles intersect? and 2) if there is a feasible solution{,} then what is the distance to voltage collapse point?. These two aspects are analyzed by representing the power flow circles as {a} family of circles{,} and this analysis is facilitated by using {the} determinant concept \cite{hans79}. 

\begin{figure}
    \centering
    \centerline{\includegraphics[scale=0.3]{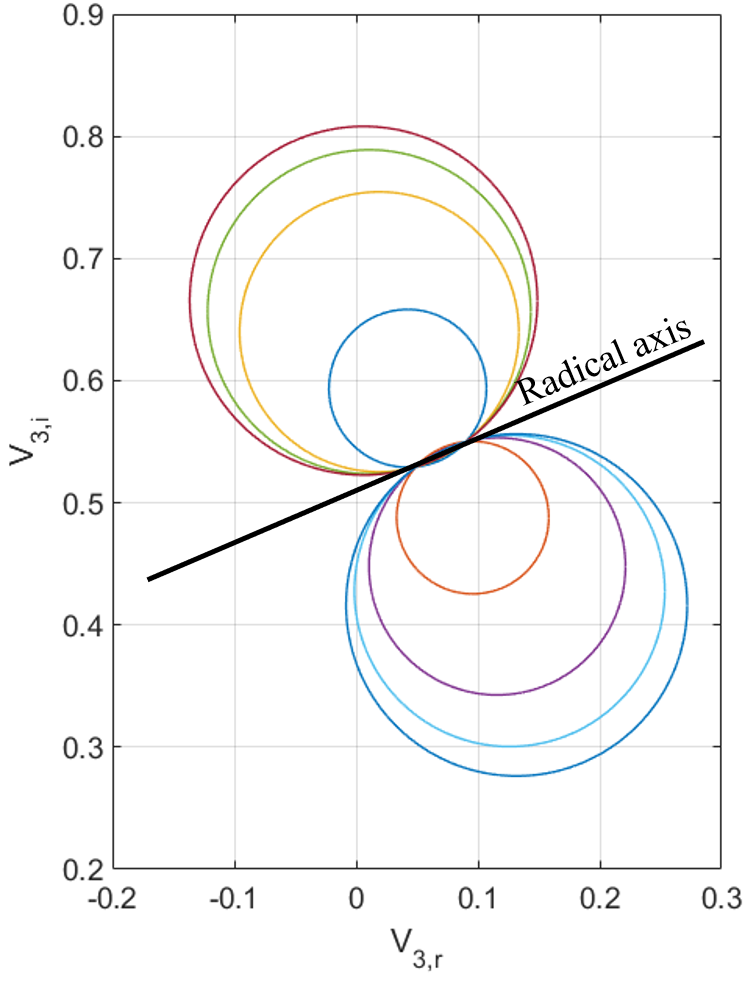}}
    \caption{Family of circles corresponding to the power flow circles (arbitrary loading condition) at bus $3$ of the $3$-bus system described in Section~\ref{subsec:impact_load_incr_on_pf_circles}.}
    \label{fig:foc3}
    \vspace{-2em}
\end{figure}

To study (1) and (2) aspects mentioned in the above paragraph, we need to analyze how the real and reactive power circles interact together i.e., family of circles (parameter of circles). Given two circles $C_1$ and $C_2$, their family of circles $C_f$ is represented by the equation $C_f = \lambda_1 \cdot C_1 + \lambda_2 \cdot C_2$ $\forall\ \lambda_1,\lambda_2\ \in \mathbb{R}$. For a given $(\lambda_1,\lambda_2)$ pair, $C_f$ represents a single circle that passes through the common points of circles $C_1$ and $C_2$. Fig.~\ref{fig:foc3} shows a typical illustration of family of circles $C_f$ with radical axis. In Fig.~\ref{fig:foc3}, the radical axis is the line that passes through the common points of circles $C_1$ and $C_2$. Table~\ref{tab:foc_radical_axis} presents some important geometrical properties of $C_f$.  
\begin{table}
  \centering
  \caption{{Properties of circles and family of circles.}}
  \includegraphics[scale=0.425]{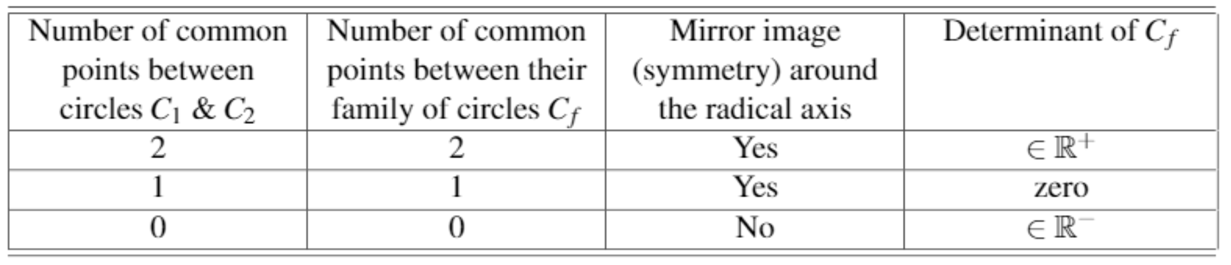}
  \label{tab:foc_radical_axis}
\end{table}

{From Tab.~\ref{tab:foc_radical_axis}, \cite{hans79} shows that the determinant can be used to study the property of symmetry of $C_f$ around the radical axis. We can see from Tab.~\ref{tab:foc_radical_axis} that the symmetry around the radical axis can directly indicate the common points between the two given circles. The following provides the inferences for the approach selected in this paper.}

{For example, when the power flow circles intersect{,} then we have common points between them and vice versa. From Table~\ref{tab:foc_radical_axis}, the mirror image (symmetry) around the radical axis is a good indicator to identify if the power flow circles intersect. Mathematically, property of symmetry around {the} radical axis in Fig.~\ref{fig:foc3} is indicated by the determinant value of $C_f$ {\cite{hans79}}. For example, a non-negative determinant value of $C_f$ indicates the symmetry around radical axis{,} i.e., power flow circles intersect{,} and thereby a feasible solution {exists}. Similarly, a negative determinant value of $C_f$ indicates that the power flow circles do not intersect.}
{In Section~\ref{subsec:derivation}, we show that family of circles $C_f$ corresponding to real and reactive power circles can be represented in a square matrix form. This square matrix form enables us to calculate its determinant.}

\subsection{Proposed Voltage Stability Indicator}
\label{subsec:derivation}
In this subsection to derive the proposed VSI using determinant, {first at given a loading condition,} the power flow circles are computed locally at a bus using PMU voltage phasor measurements from its adjacent buses and these circles are then represented as square matrices to enable the determinant calculation. Second, we derive the proposed VSI by utilizing the properties of determinant inferred in Section~\ref{subsec:quant_dist_btwn_pf_circles}.

Equation \eqref{eq:comp_eq} represents the standard homogeneous form of a circle ({$C_k$}) with locus $z=x+j\cdot y$ and $\mathit{z^*}$ is complex conjugate of $z$. {The center and radius of the circle $C_k$ are denoted by $\gamma=\alpha + j \cdot \beta$ and $\rho$ respectively.} Reference \cite{hans79} shows the idea of representing circle as a hermitian matrix as shown in \eqref{eq:herm_rep}. {The derivation from \eqref{eq:comp_eq} to \eqref{eq:herm_rep} is provided in Appendix~\ref{subsec:app:mat_rep_std_circles}.}
\begin{align}
\mathit{{C_k}\left( z, z^* \right)} &= \mathit{A} \cdot \mathit{{z \cdot z^*}} + \mathit{B} \cdot \mathit{{z}} + \mathit{C} \cdot \mathit{{z^*}} + D = 0, \label{eq:comp_eq} \\
&\triangleq \begin{bmatrix} A & B \\ C & D \end{bmatrix}\ \triangleq  \begin{bmatrix} 1 & -\gamma^* \\ -\gamma & (\gamma\cdot\gamma^* - \rho^2) \end{bmatrix}, \label{eq:herm_rep}
\end{align}
where $\mathit{A} \neq 0$. $\mathit{A}$ and $\mathit{D}$ are always real, $\mathit{B}$ and $\mathit{C}$ are always complex conjugates for a circle. Since the power flow equations are also circles, they can be represented as hermitian matrices and such matrix representation of power flow equations enable the determinant calculation which helps to derive the proposed index.

\textbf{Matrix Representation of Power Flow Equations and Proposed VSI:}
Once {we obtain the matrix form of power flow equations}, its determinant is useful to 1) identify the existence of power flow solution and 2) distance to voltage collapse point. Similar to the above calculations, the elements of real power circle-matrix $\mathcal{C}_p$ are calculated using its center \eqref{eq:cpq} and radius \eqref{eq:rp} as shown below. 
\begin{align}
    &\mathcal{C}_p \triangleq \begin{bmatrix} A_p & B_p \\ C_p & D_p \end{bmatrix}, \notag \\
    &\mathit{A}_{p} = 1, \mathit{B}_{p} = -\left({o_p}\right)^* = -\left(\dfrac{-\bd b_{p}}{2}\right)^* = \left(\dfrac{\bd b_{p}}{2}\right)^*, \notag \\
    &\mathit{C}_{p} = -\left({o_p}\right) = -\left(\dfrac{-\bd b_{p}}{2}\right) = \left(\dfrac{\bd b_{p}}{2}\right), \notag \\
    &\mathit{D}_{p} = \left({o_p}\right)\cdot\left({o_p}\right)^* - \mathit{r}^2_{p} = \mathit{c}_{p},
\end{align}
where $\bd b_{p} = \begin{bmatrix}\dfrac{t_{d,2}}{t_{d,1}} & \dfrac{t_{d,3}}{t_{d,1}}\end{bmatrix}^{T}$, $\bd b_{q} = \begin{bmatrix} \dfrac{-t_{d,3}}{t_{d,4}} &  \dfrac{t_{d,2}}{t_{d,4}}\end{bmatrix}^{T}$, $\mathit{c}_p = \dfrac{-\mathit{p}_{d}}{t_{d,1}}$ and $\mathit{c}_q = \dfrac{-\mathit{q}_{d}}{t_{d,4}}$. Similarly, the reactive power circle-matrix $\mathcal{C}_q$ is also shown below.
\begin{align}
\mathcal{C}_p &= \begin{bmatrix}1 & \left(\dfrac{\bd b_{p}}{2}\right)^* \\ \left(\dfrac{\bd b_{p}}{2}\right) & \mathit{c}_{p}\end{bmatrix},
\mathcal{C}_q = \begin{bmatrix}1 & \left(\dfrac{\bd b_{q}}{2}\right)^* \\ \left(\dfrac{\bd b_{q}}{2}\right) & \mathit{c}_{q}\end{bmatrix}. \notag
\end{align}
Finally, the power flow equations are represented as matrices $\mathcal{C}_p$ and $\mathcal{C}_q$ respectively. {As discussed in Section~\ref{subsec:quant_dist_btwn_pf_circles}, to study the voltage collapse phenomenon one must analyze the real and reactive power circles together i.e., family of circles ($\lambda_1\cdot \mathcal{C}_p + \lambda_2\cdot \mathcal{C}_q$). Using the inferences and property of determinant described in Section.~\ref{subsec:quant_dist_btwn_pf_circles}, at a given loading condition $C_f$ is given by $\mathcal{C}_p + \mathcal{C}_q$ and the expression for its determinant $\left|C_f\right|$ (say $\Delta^*$) is given by \eqref{eq:ref_me}.}

\begin{align}
	\Delta^{*} &= \Delta_p \cdot \Delta_q - \Delta^2_{pq}. \label{eq:ref_me}
\end{align}

where
\begin{align}
	\Delta_{p} &= \left(\mathit{c}_p - \dfrac{\left\|\bd b_{p}\right\|^2}{4} \right), 
	\Delta_{q} = \left(\mathit{c}_q - \dfrac{\left\|\bd b_{q}\right\|^2}{4} \right), \notag \\
	\Delta_{pq}&= \left(\dfrac{1}{8} \cdot \left\| \bd b_{p} - \bd b_{q} \right \|^2 - \dfrac{1}{2} \cdot \left(\dfrac{\left\|\bd b_{p}\right\|^2}{4} - \mathit{c}_p \right) - \right. \notag \\ &\left. \dfrac{1}{2} \cdot \left(\dfrac{\left\|\bd b_{q}\right\|^2}{4} - \mathit{c}_q \right)\right)^2. \notag
\end{align}

{We determine $\Delta^{*}$ by calculating $\Delta_p$, $\Delta_q$ and $\Delta_{pq}$ at a given loading condition using the centers and radii.} The power flow circles touch each other at the voltage collapse point (nose point of {the} PV curve). In such a scenario, Tab.~\ref{tab:foc_radical_axis} shows that the determinant $\Delta^*$ should become zero. Hence the lower bound of $\Delta^{*}$ is zero{,} which occurs at {the} nose point of {the} PV curve (externally touching circles). However, to interpret the results of the proposed index{,} it should range between $1$ and $0$. In order to bound the upper limit of $\Delta^{*}$, we normalize it using the no-load value of the index. The normalized form of $\Delta^{*}$ is the proposed VSI{,} as shown below.
\begin{align}
\text{Proposed VSI} \equiv  \Delta^*_{\textrm{norm}} = \dfrac{\Delta^*}{\Delta^*_{\textrm{no-load}}},\label{eq:vsi_proposed}
\end{align}
where $\Delta^*_{\textrm{no-load}}$ is the value of \eqref{eq:ref_me} with zero load. Appendix~\ref{subsec:app:proof_linearity} provides an explanation of the near-linear behaviour of the proposed index for a $2$-bus system.
\begin{remark}
    (\textbf{VSI for PV Buses}): \eqref{eq:vsi_proposed} is the distributed VSI for a PQ bus{,} and a similar VSI can be derived for a PV bus. For PV buses, the voltage solution is determined by the intersection of voltage and real power circles{,} and the distance between them indicates the VAR limit violation. These circles are again computed using the PMU voltage phasor measurements of adjacent buses.
\end{remark}  
	%--
	\section{Simulations and Discussion}
	\label{sec:Simulations}
	The proposed distributed non-iterative VSI is tested on various test cases such as IEEE $30$, $300$ and $2383$-bus systems \cite{matfile}. To obtain the voltage phasor measurements and true voltage stability margin, {we use an adaptive Newton-Raphson step size based power flow solver known as continuation power flow ({CPF}) \cite{ajjarapu1994} from MATPOWER \cite{ZimmermanEtAl2011}. CPF calculates the voltage solution on the PV curve by increasing the load and generation along a load/generator increase direction parametrized by continuation parameter $\lambda$ \cite{ajjarapu1994}.}

In this section, we show that the proposed distributed VSI is advantageous compared to both centralized and decentralized VCPIs. We also compare the proposed distributed index with the only other distributed method \cite{Porco16} in the literature to show its merits. Finally, we show the behavior of {the} proposed index for line \& generator outages.

\subsection{Non-iterative Distributed VSI Versus Centralized VCPIs}
\begin{table}[H]
	\centering
	\caption{Comparison between centralized, proposed distributed VSI and decentralized VCPIs to monitor buses $14,29$ and $30$ in IEEE-$30$ bus system \cite{matfile}.
	}
	\renewcommand{\arraystretch}{1.5}
	\begin{tabularx}{\columnwidth}{|>{\centering} p{2.25cm}|>{\centering} p{1cm}|>{\centering} p{1cm}|>{\centering} p{1.5cm}|p{0.9cm}|}
		\hline \hline
		VCPI calculation method & No. of PMUs required & Sensitivity to noise & Full admittance matrix & Single point failure\\ \hline
		Centralized  & $30$ & Low & Required & Yes \\ \hline
		{Distributed : iterative index \cite{Porco16}} & {$30$} & {Low} & {Not required} & {No} \\ \hline
		Distributed : proposed index & $5$ & Low & Not required & No \\ \hline 
		Decentralized  & $3$ & High & Not required & No\\ \hline \hline
	\end{tabularx}
	\label{tab:dist_comp_centr}
	\vspace{-1em}
\end{table}

The power system operator generally monitors the power system from a control room. In practice, there are specific regions of the power system that are critical for LTVI and can be identified by offline means. Thus, the operator is interested in online LTVI monitoring in a few strategic locations/regions. For instance, in the IEEE-$30$ bus system, the operator would like to monitor the buses $14,29$ and $30$. To calculate centralized VCPIs at buses $14,29$ and $30$, the voltage measurements at all the buses in the power system are required{,} i.e., full observability. To calculate the proposed distributed index at buses $14,29$ and $30$, the voltage measurements at only their immediate neighboring buses are required. The adjacent nodes connected to buses $14$, $29$ and $30$ are buses $(12, 15)$, $(27,30)$ and $(27,29)$ respectively. Thus we only require $5$ PMUs at buses $12,15,27,29$ and $30$ to calculate the proposed distributed VSI. 

Additionally, the proposed VSI does not require the complete knowledge of {the} system admittance matrix. Instead{,} it takes advantage of the sparse nature of power system graph by using only branch admittance values of branches connecting to buses $14,29$ and $30$ to calculate the proposed distributed index. Table~\ref{tab:dist_comp_centr} presents the requirements of centralized, decentralized, and distributed methods such as proposed index, existing iterative index \cite{Porco16} to monitor buses $14,29$ and $30$. {It is important to note that even though the iterative distributed index proposed in \cite{Porco16} uses communication links between the neighboring buses, \cite{Porco16} mandates requirement of PMUs at all buses in the grid for convergence of its iterative algorithm. {Whereas} the proposed method does not require PMUs to cover all the buses.}

\subsection{Non-iterative Distributed VSI Versus Decentralized VCPIs}
\begin{figure}
    \centering
    \centerline{\includegraphics[scale=0.25]{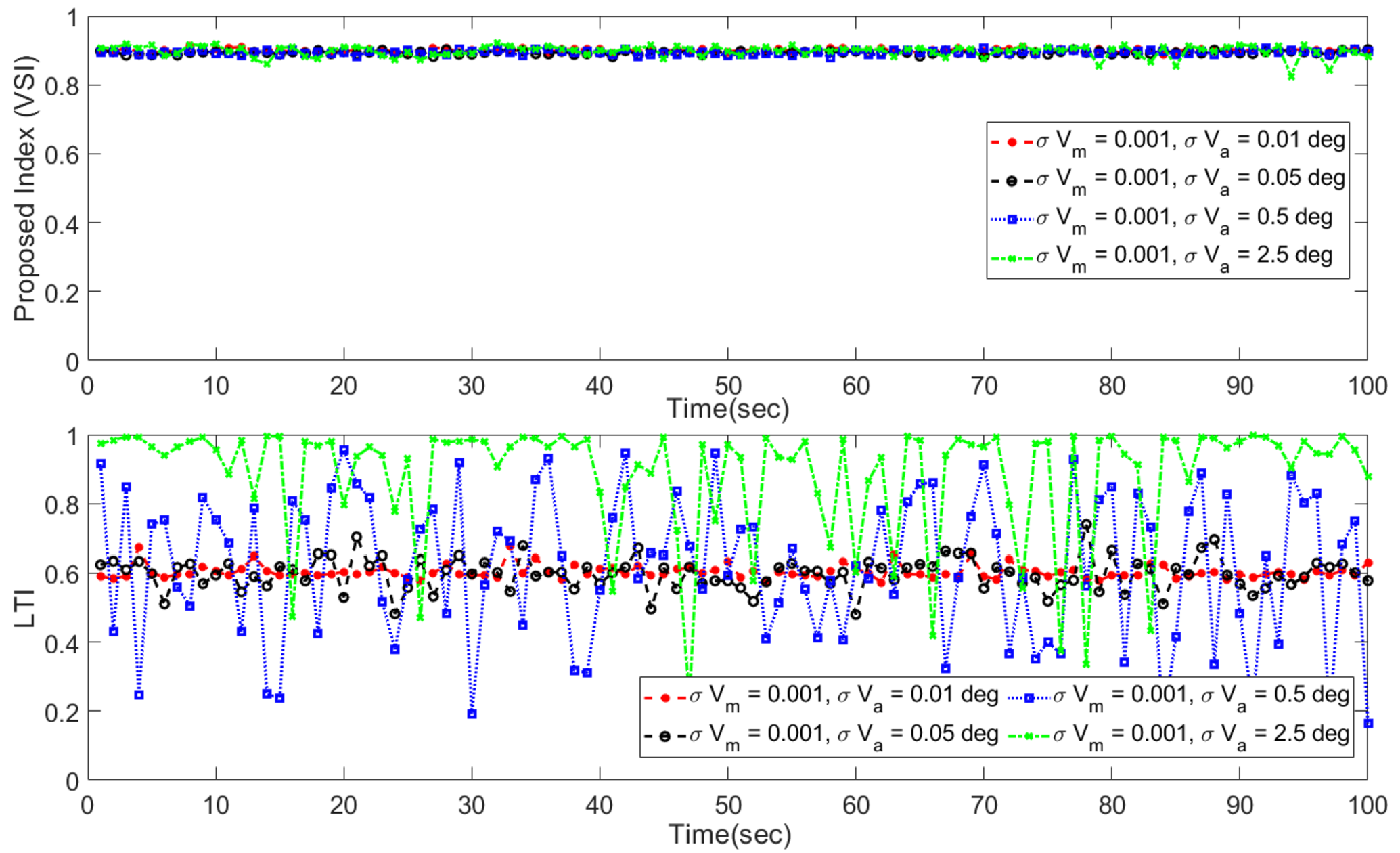}}
    \caption{Effect of noisy measurements with different noise levels on decentralized (LTI) and proposed distributed index. LTI is more sensitive to noise due to its approximation errors.}
    \label{fig:noise_comparison_plot}
    \vspace{-2em}
\end{figure}

In this subsection, we compared the proposed VSI with other methods (decentralized and distributed index \cite{Porco16}) when there is noise in the PMU measurements{,} and we show that the proposed VSI is less prone to noisy measurements. {To understand the impact of noise on the proposed methodology, an additive Gaussian noise with zero mean and standard deviation of $0.001$ p.u. on voltage magnitude ($\sigma V_m=0.001 p.u.$), $0.5^\circ$ on phase angles ($\sigma V_a = 0.5^{\circ}$) are introduced in the measurements according to the analysis of field-tested PMUs by New England ISO \cite{zhang2013,brown16} and IEEE standard for acceptable PMU errors \cite{martin15}.} To demonstrate the robustness of the proposed VSI with regard to noise, it is compared to the local Thevenin index (LTI) i.e., a decentralized VCPI \cite{Vu99,Verbic04,Corsi08} and distributed sensitivity index \cite{Porco16}. {The results are shown in  Tab.~\ref{tab:noise_table} and Fig.~\ref{fig:noise_comparison_plot} under different noise levels}.
\begin{table}[!htbp]
	\centering
	\caption{The standard deviation ($\sigma$) for the proposed VSI, LTI and distributed sensitivity index \cite{Porco16} at bus $30$ in IEEE-$30$ bus system when Gaussian noise %with zero mean and variable standard deviations $\sigma V_m$ and $\sigma V_a$ 
	is introduced in the voltage phasor measurements.}
	\renewcommand{\arraystretch}{1}
	\begin{tabularx}{\columnwidth}{|P{2.3cm}|P{1.3cm}|P{1.3cm}|P{2.2cm}|}
		\hline \hline
		Noise in voltage angle ($\sigma V_a$) and magnitude ($\sigma V_m$ ) & Proposed index (VSI)~~~~~& Local Thevenin index (LTI) &Distributed iterative sensitivity index \cite{Porco16}\\ \hline
		$\sigma V_m=0.001 p.u.$ $\sigma V_a=0.01^{\circ}$, \cite{martin15} & $\sigma = 0.0043$ & $\sigma = 0.0184$ & $\sigma = 0.0057$ when convergence tolerance = $0.001$ \\ \hline
		$\sigma V_m=0.001 p.u.$ $\sigma V_a=0.5^{\circ}$, \cite{zhang2013} & $\sigma = 0.0058$ & $\sigma = 0.1942$ & $\sigma = 0.0084$ when convergence tolerance = $0.1$\\ \hline
		$\sigma V_m=0.001 p.u.$ $\sigma V_a=0.5^{\circ}$, \cite{zhang2013} & $\sigma = 0.0058$ & $\sigma = 0.1942$ & algorithm diverges when convergence tolerance = $0.001$\\ \hline\hline
	\end{tabularx}
	\label{tab:noise_table}
	\vspace{-1em}
\end{table}
% \vspace{-2.5em}

{From Tab.~\ref{tab:noise_table}, it can be observed that the proposed VSI has {a} minimal standard deviation when compared to that of both LTI and distributed sensitivity index \cite{Porco16} for different noise levels. Additionally, the standard deviation of LTI increases considerably when $\sigma V_a = 0.5^{\circ}$. However, the proposed index and distributed iterative sensitivity index \cite{Porco16} are not so impacted by the increase in the noise of voltage angle measurements.} {Despite the lower standard deviation of distributed iterative sensitivity index in Tab.~\ref{tab:noise_table}, it is important to note that the method \cite{Porco16} suffers from non-convergence as the noise increases. For example, when $\sigma V_m=0.001 p.u.$  and $\sigma V_a=0.5^{\circ}$, the distributed sensitivity index \cite{Porco16} does not converge when its convergence tolerance is $0.001$ but converges for a lower level noise of $\sigma V_a = 0.01^\circ$ with same tolerance setting. This creates a challenge in setting the threshold for convergence \cite{Porco16} when deployed in the field since the PMU measurements have variable noise levels. In contrast, the proposed index has no such drawback as it is a non-iterative index and it is also least sensitive to the noise in PMU measurements as shown in Tab.~\ref{tab:noise_table}.} 

From Fig.~\ref{fig:noise_comparison_plot}, LTI is more sensitive to noise than that of the proposed distributed index due to its approximation errors \cite{Vu99}. Similarly, from Tab.~\ref{tab:noise_table}, the distributed iterative index is more sensitive to noise than the proposed distributed index due to its iterative nature. While the proposed distributed index is the least sensitive one to noise.

\subsection{Effect of PMU Voltage Phasor Data: Measurement Noise Variability, Large Measurement Outliers{,} and Missing Data}
\begin{figure}
    \centering
    \includegraphics[scale=0.25]{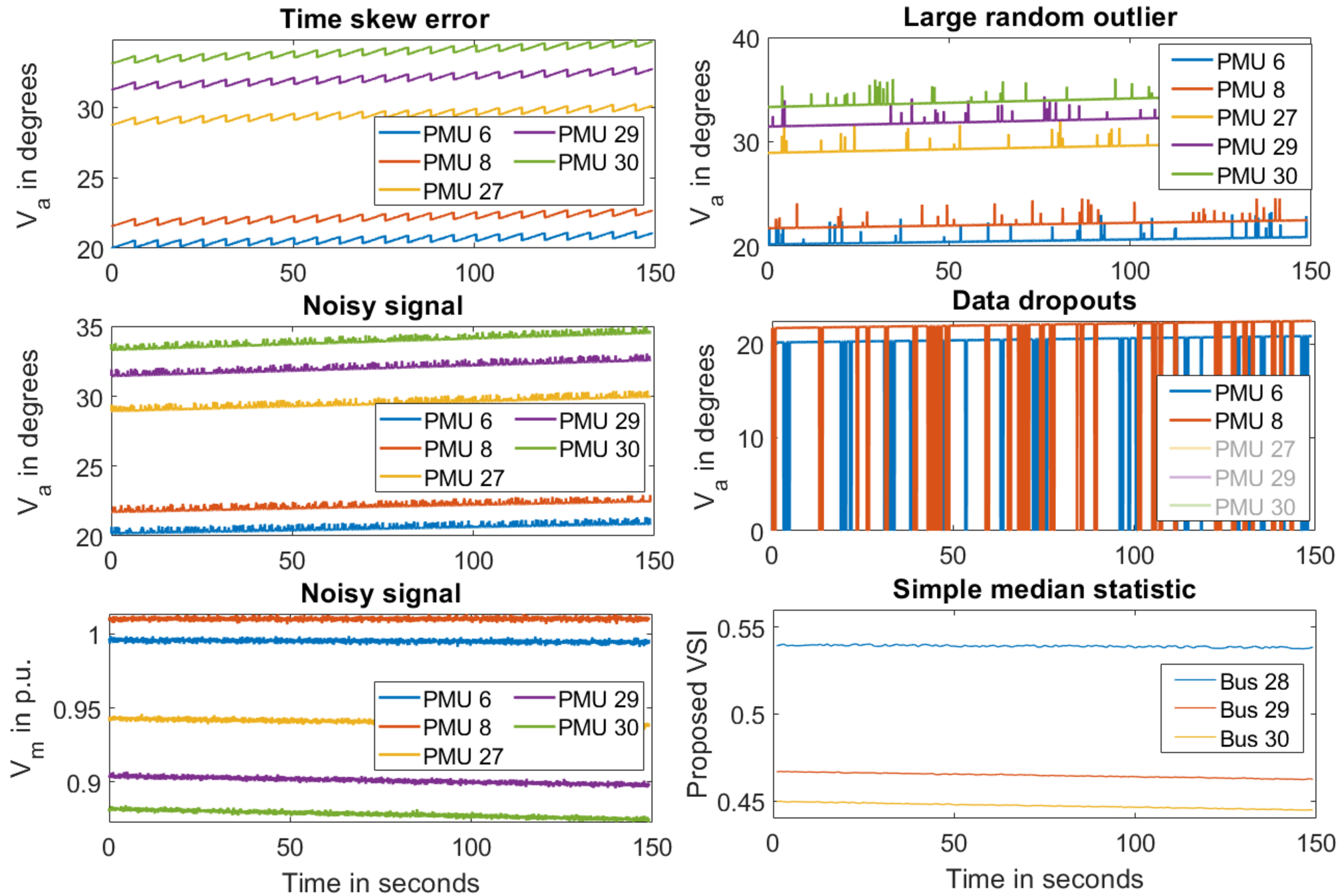}
    \caption{The subplots in this figure present the different types of PMU errors observed in real-world \cite{overbye19}. These different noisy measurements are generated by using the error statistics from \cite{overbye19}. %The different noises include time skew errors, large random outliers, noisy signals and data drop outs. 
    The last subplot shows the performance of the proposed index that uses a simple median statistic on a one second window size to handle various bad quality PMU data mentioned above.}
    \label{fig:dsdsds}
    \vspace{-1em}
\end{figure}

We further investigate the effect of bad quality phasor measurements from PMUs on the proposed index. To handle these bad quality phasor measurements, one must use a preprocessor that takes bad quality measurements like input and outputs the filtered/better measurements. These filtered measurements are used to calculate the proposed distributed index. Depending on the quality of data, this preprocessor can be a state estimator, bad data detector, low-rank matrix methods, etc. and this preprocessor is independent of the proposed methodology. Various types of errors such as clock drift errors, measurement spikes, high amount of noise{,} and missing data can lower the quality of the PMU data \cite{overbye19}. We generate PMU voltage phasor data that is very close to the real-world using the error statistics presented in \cite{overbye19,zhang2013} as shown in Fig.~\ref{fig:dsdsds}. In this subsection, we use noisy PMU measurements to calculate the proposed index at buses $28,29,$ and $30$ of IEEE-$30$ bus system. Since these noisy measurements with different types of errors can directly impact the quality of the VCPIs, the proposed index is calculated by using the output measurements of a preprocessor that uses simple robust statistic such as median of the PMU measurements in {the} given time window (e.g., 1 sec). Fig.~\ref{fig:dsdsds} shows that the proposed index is not affected by any of the data dropouts, measurement spikes{,} and the device clock errors. In contrast to \cite{Porco16}, due to the iterations involved, it is unclear how the distributed iterative index \cite{Porco16} handles the different kinds of noise like missing data, time skew, measurement spikes and large white noise. Different methodology based preprocessors provide different quality of measurements. Irrespective of the preprocessor, for a given same amount of input noise signal, we show that the proposed index is robust (Tab.~\ref{tab:noise_table}, Fig.~\ref{fig:noise_comparison_plot}) and fast (Tab.~\ref{tab:time_iter_porcos_vsi_compare}, Tab.~\ref{tab:flops_comp}) when compared to other methods \cite{Vu99,Verbic04,Corsi08,Porco16}.

\begin{figure}
    \centering
    \includegraphics[scale=0.25]{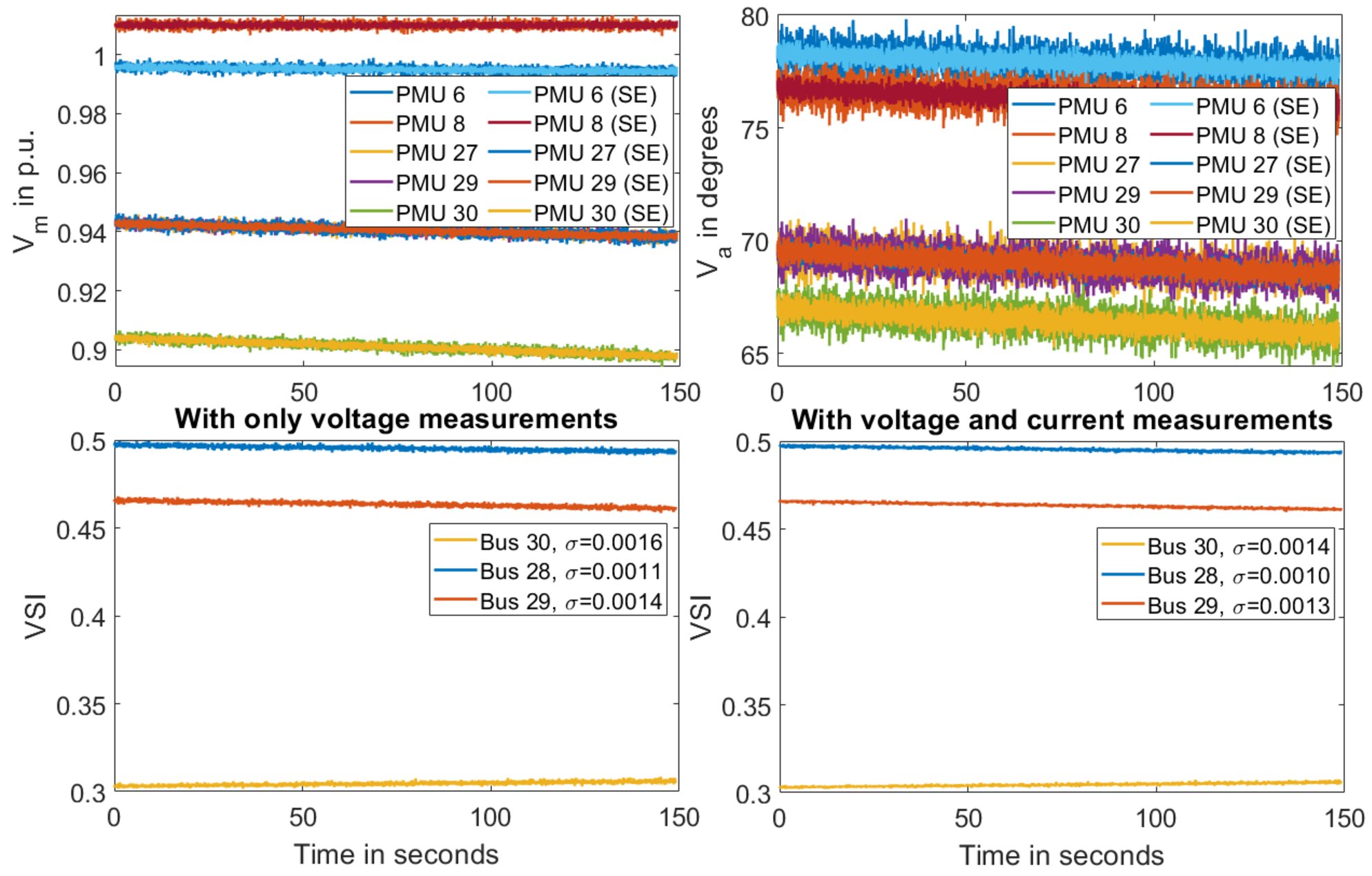}
    \caption{Preprocessor 1: Reduction of noise in PMU measurements using only voltage or both voltage and current measurements (SE). Improvement in the accuracy of proposed VSI due to better quality measurements.
    }
    \label{fig:preprocessor_se}
    \vspace{-1em}
\end{figure}

\begin{figure}
    \centering
    \includegraphics[scale=0.25]{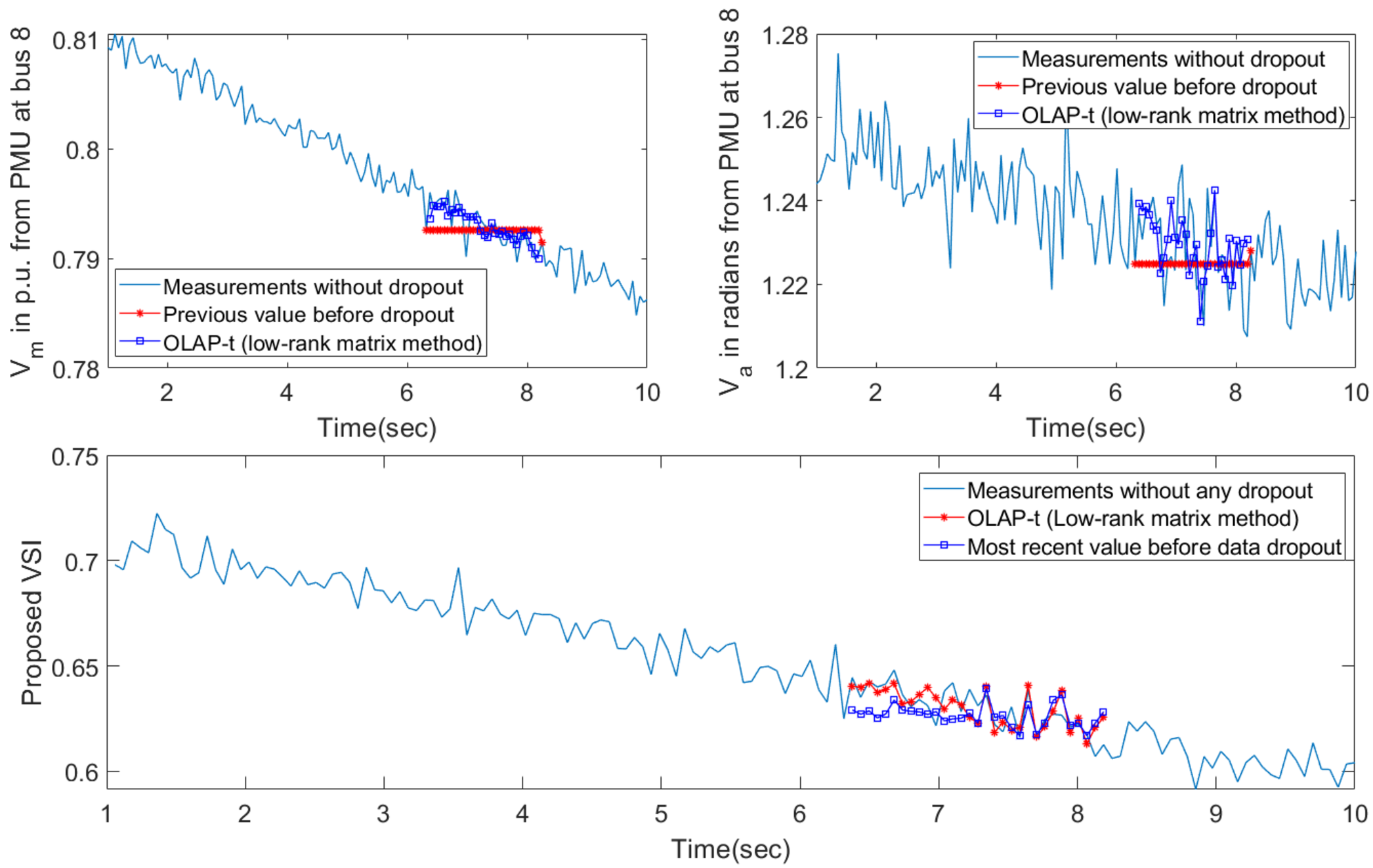}
    \caption{Preprocessor 2: Recovery of large chunks of missing data using \cite{de2018evaluation} and a simple heuristic that uses the last known non-zero measurement value. The output of the preprocessor with \cite{de2018evaluation} improves the accuracy of the proposed index when there are large chunks of missing data from PMUs due to their CT/PT failures.}
    \label{fig:preprocessor_lowrank_method}
    \vspace{-2em}
\end{figure}

Thus, the proposed methodology is robust to various kinds of noise types and levels when compared to \cite{Vu99,Porco16,Verbic04,Corsi08}, ensuring that the alarms triggered using this VSI will have fewer false-positive rates, providing a reliable grid monitoring scheme.
% \newpage
For an illustration of the impact of a preprocessor on the proposed index, we present two cases i.e., 1) distributed state estimator that uses both current and voltage measurements, and 2) low-rank matrix method. 1) Fig.~\ref{fig:preprocessor_se} shows a preprocessor that uses the same distributed communication architecture requirement as that of the proposed VSI can use redundant local branch current measurements to improve the noisy data and thereby increase the accuracy of the proposed method. Generally, the proposed index is robust to noise. However, in cases when there is an impact of noise, it is recommended to use a distributed state estimation based preprocessor that uses redundant branch current measurements to further improve the performance of the proposed methodology. 2) Fig.~\ref{fig:preprocessor_lowrank_method} shows a preprocessor that can recover large chunks of missing data where a simple median statistic will fail. Specifically, we used the OLAP-t method \cite{de2018evaluation} to recover the missing data in a $2$ second window frame from a PMU located at bus $8$. In the IEEE-30 bus system, buses $6,8,27$ are the neighboring buses to bus $28$. It is important to use data only from buses $6,8,27$ to recover the missing data at bus $8$ due to the distributed communication architecture (framework) used by the proposed methodology. We used the recovered data from \cite{de2018evaluation} and a simple heuristic that uses last known non-zero values to calculate the proposed index at bus $28$. Fig.~\ref{fig:preprocessor_lowrank_method} shows that \cite{de2018evaluation} is more accurate than the simple heuristic method especially when there is more missing data.

\subsection{Non-iterative Distributed VSI Versus Iterative Distributed Sensitivity Index}
In this subsection, we compare the interpretability of the proposed index and iterative distributed sensitivity index \cite{Porco16} and show that the proposed index can provide a measure of distance to system voltage collapse. We also discuss the number of iterations taken by \cite{Porco16} in contrast with the non-iterative nature of the proposed index. 
\subsubsection{Proportional Load \& Generation Increase}
\begin{figure}
\centering
\begin{subfigure}{\columnwidth}
    \centerline{\includegraphics[height=3.9cm, width=\columnwidth]{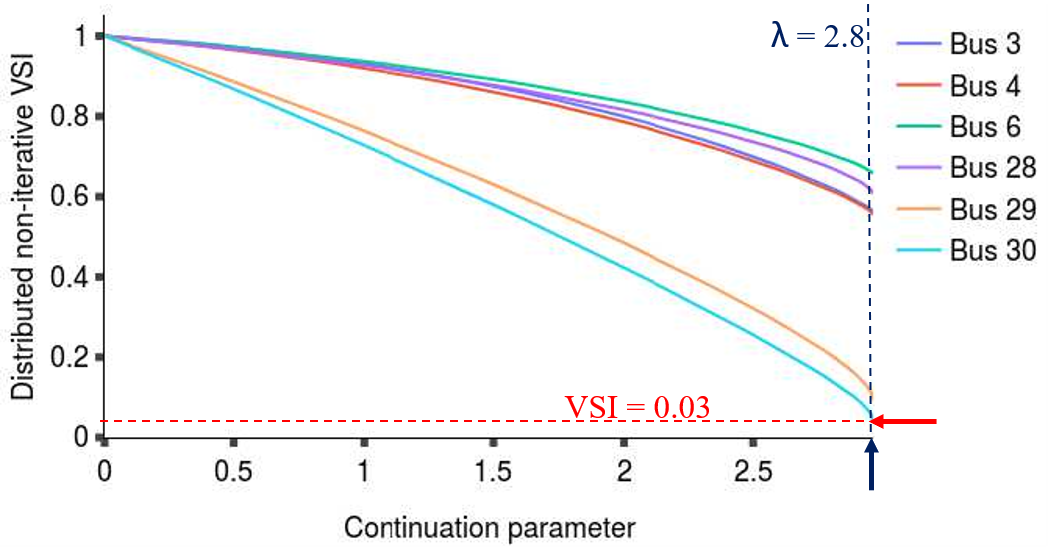}}%Case30_wo_Qlimits.eps
    \caption{Proposed index values with increase in system loading to a critical value ($\lambda=2.8$) from no load condition ($\lambda = 0$).}
    \label{fig:30-bus without Q}
\end{subfigure}
\begin{subfigure}{\columnwidth}
    \centerline{\includegraphics[height=3.99cm, width=\columnwidth]{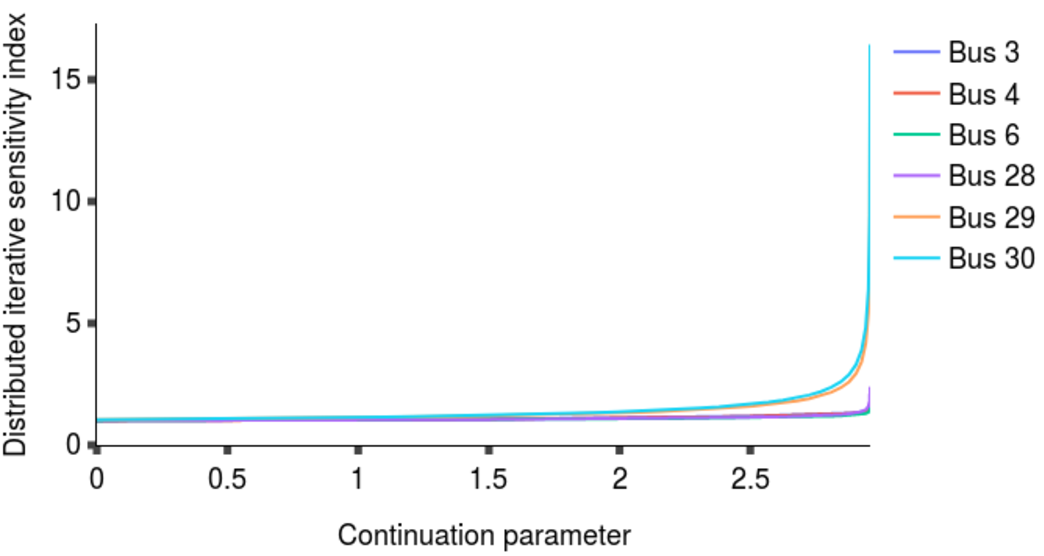}}%porco_Case30_wo_Qlimits.eps
    \caption{Index values from \cite{Porco16} with increase in system loading to a critical value ($\lambda=2.8$) from no load condition ($\lambda = 0$).}
    \label{fig:30-bus without Q_porco}
\end{subfigure}
\caption{Index values of proposed VSI and \cite{Porco16} versus system load scaling factor.}
\vspace{-2em}
\end{figure}

In this case, first, we verify the proposed VSI correctly identifies the critical bus in the system that causes the LTVI by comparing it with the only other distributed method \cite{Porco16} in the literature. Second, we show that the proposed VSI is easier to interpret compared to \cite{Porco16}. {$\lambda$ (continuation parameter from CPF) is defined as the load scaling factor that scales the loads and generations in IEEE-$30$ bus system} and {continuation power flow (CPF) solver \cite{ajjarapu1994} from MATPOWER \cite{ZimmermanEtAl2011} is used to generate voltage phasor measurements for each $\lambda$}. The voltage measurements for each $\lambda$ are used to calculate the proposed non-iterative distributed VSI. Fig.~\ref{fig:30-bus without Q} shows the proposed distributed index values versus load scaling factor $\lambda$ for IEEE-$30$ bus system. The first observation is that at the critical loading (corresponding to $\lambda = 2.8$), bus $30$ has the lowest voltage magnitude of $0.51$ p.u. in the entire network. The reason for bus $30$ to be critical is due to its location being electrically farthest away from the generators and synchronous condensers. When the load scaling factor $\lambda$ is greater than $2.8$, there is no feasible operating solution identified by CPF due to the LTVI phenomenon. Consequently, when $\lambda = 2.8$, the proposed distributed index value calculated at bus $30$ is the smallest among all buses, and it is very close to 0, implying that this is the critical bus in the system that causes the LTVI. 

\begin{figure}
\centering
\begin{subfigure}{\columnwidth}
    \centerline{\includegraphics[scale=0.35]{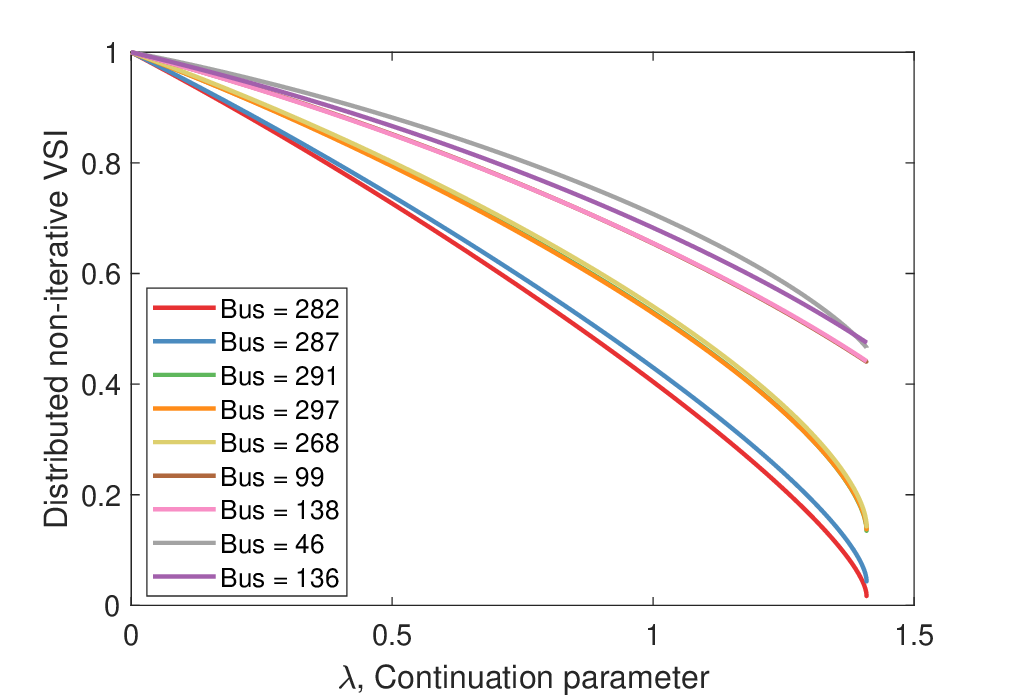}}%Case30_wo_Qlimits.eps
    \caption{The distributed non-iterative voltage stability index for IEEE-$300$ bus systems. The weakest bus when the entire load in the system proportionally increases is bus $282$.}
    \label{fig:300_pic}
\end{subfigure}
\begin{subfigure}{\columnwidth}
    \centerline{\includegraphics[scale=0.35]{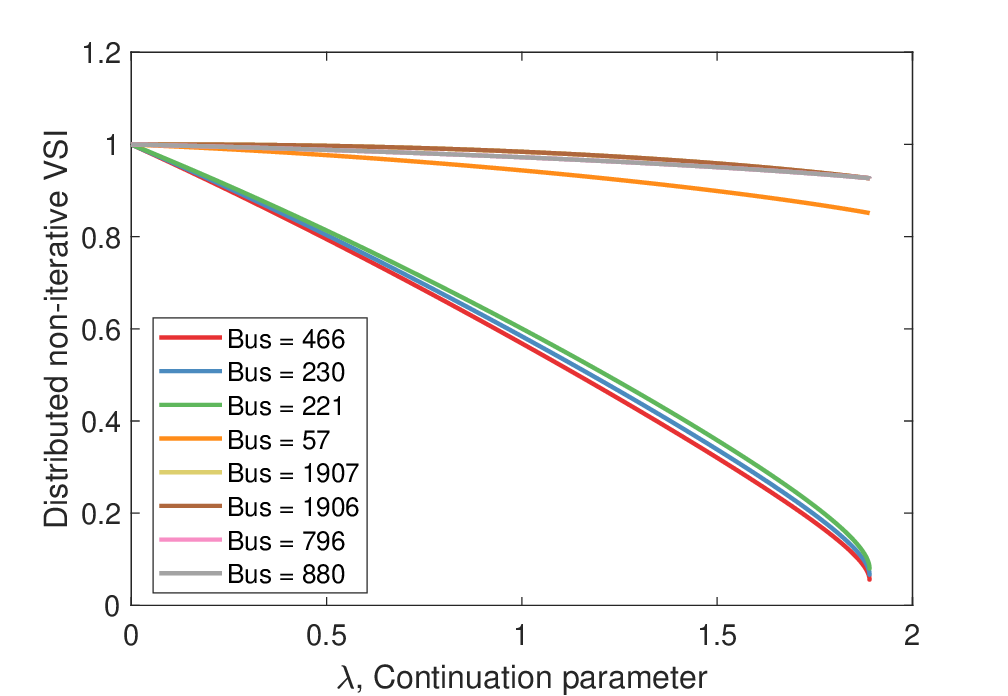}}%porco_Case30_wo_Qlimits.eps
    \caption{The distributed non-iterative voltage stability index for IEEE-$2383$ bus systems. The weakest bus when the entire load in the system proportionally increases is bus $466$.}
    \label{fig:2383_pic}
\end{subfigure}
\caption{The distributed non-iterative voltage stability index for IEEE-$300$ and IEEE-$2383$ bus systems.}
\vspace{-2em}
\end{figure}

This fact is also verified by comparing {it} with {a} distributed iterative sensitivity method \cite{Porco16}. Fig.~\ref{fig:30-bus without Q_porco} presents the index values of \cite{Porco16} versus load scaling factor ($\lambda$) and it also shows that the critical bus is bus $30$ i.e., the bus with maximum sensitivity index value at $\lambda = 2.8$. Thus the VSI correctly identifies the critical bus in the system. However, It can be seen that the sensitivity index varies in an extremely non-linear manner with the load scaling (continuation) parameter ($\lambda$). Distributed iterative sensitivity index in Fig.~\ref{fig:30-bus without Q_porco} is unbounded at $\lambda=2.8$, making it hard to interpret the distance to voltage collapse point. Whereas, the distributed non-iterative VSI in Fig.~\ref{fig:30-bus without Q} is bounded between $1$ and $0$ corresponding to no-load and loadability limit, respectively. This makes the proposed index a good indicator to interpret the distance to voltage collapse point. Hence the non-interpretability and non-linear nature of the sensitivity methods makes it hard to set monitoring thresholds to reliably trigger controls while the proposed index solves these problems by effectively using the PMU measurements. {A similar study for larger test case systems such as IEEE-$300$ and IEEE-$2383$ along with their critical buses are also presented in Fig.~\ref{fig:300_pic} and Fig.~\ref{fig:2383_pic} respectively.}
% \newpage

{Even though the proposed index value at the critical bus (e.g., bus $30$) is expected to be $0$ at critical loading ($\lambda = 2.8$), we observe that the index value is a very small non-zero value (VSI = $0.03$) from Fig.~\ref{fig:30-bus without Q}. This behavior is due to the numerical instability (ill-conditioned Jacobian) of the power flow solver (CPF) from MATPOWER \cite{ZimmermanEtAl2011}, which is used to generate the voltage phasor measurements. For example, when the loading condition $\lambda$ is very close to the nose point of the PV curve, it is well-known that the power flow Jacobian becomes ill-conditioned and power flow solvers diverge \cite{Iwamoto81}. Due to this divergence behavior, it is not possible to accurately generate the voltage phasor measurements corresponding to the nose point of the PV curve, where the proposed index becomes zero. Hence we see very small non-zero values in Fig.~\ref{fig:30-bus without Q}. However, this limitation is not present when the proposed index is deployed on the field because the voltage phasor measurements are taken from the PMUs directly.}

\subsubsection{Computation Performance of VSI}
In this case, we first compare the performance of the proposed distributed non-iterative index with that of the distributed iterative index \cite{Porco16} under two scenarios. They are a) increase in system loading and b) the presence of noise in voltage phasor measurements. a) In case of an increase in system loading $\lambda$, from Tab.~\ref{tab:time_iter_porcos_vsi_compare}, in the presence of both noise and no noise, the time is taken by the distributed iterative method \cite{Porco16} increases with increase in system loading $\lambda$. This behavior is not desirable since the updating frequency (calculation time) of the distributed iterative index is not constant, and it is dependent on the loading condition ($\lambda$). Fortunately, the time taken by the proposed index is constant and almost negligible for any loading $\lambda$ due to its time complexity being $O(1)$ (non-iterative index). b) In presence of noise in voltage phasor measurements, it can be seen from Tab.~\ref{tab:time_iter_porcos_vsi_compare} that for any system loading $\lambda$, the time taken by the distributed iterative index \cite{Porco16} is higher than that of the no noise scenario. To speed-up the distributed iterative index, its convergence tolerance has to be varied based on the noise level in measurements. However, it is not practical to adjust the convergence tolerance depending on varying noise levels in an online monitoring application. Fortunately, again the time taken by the proposed method is not affected by the noise level in measurements due to its approximation free formulation and non-iterative nature.

\begin{table}
  \centering
  \caption{{Total time and iterations taken by the proposed and existing method \cite{Porco16} with different directions of load increase at buses $1-16$ and $17-30$ for various system loading values ($\lambda$) and noise.
  }}
  \includegraphics[scale=0.4]{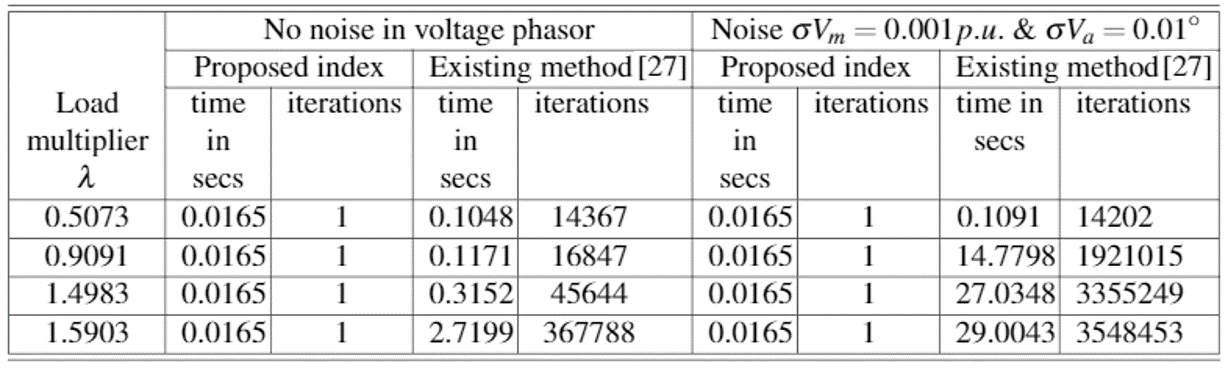}
  \label{tab:time_iter_porcos_vsi_compare}
  \vspace{-2em}
\end{table}

{For a better comparison of the proposed and existing iterative method \cite{Porco16}, Tab.~\ref{tab:flops_comp} presents the totals floating-point operations per second (Flops) taken by both the methods. To calculate the flops, we use the information from \cite{thant05} to determine the count of flop needed for performing mathematical operations such as addition, subtraction, multiplication, and division. The number of flops required to calculate the proposed index at bus $n$, which has ``$M$" neighboring buses = $8 \cdot M + 37$. While, the total count of flop required to calculate the existing iterative method = $16 \cdot M + (16 \cdot M +360) \cdot I$, where ``$I$" is the total iterations taken by the iterative method to converge. In case of the proposed method, since it is a non-iterative approach the required number of flops expression does not have ``$I$" (number of iteration) in it.}
\begin{table}
  \centering
    \caption{{Total time, iterations, Flop and Flops (flop per second) taken by the proposed and existing \cite{Porco16} methods to calculate their indices value for bus $30$.
    }}
  \includegraphics[scale=0.4]{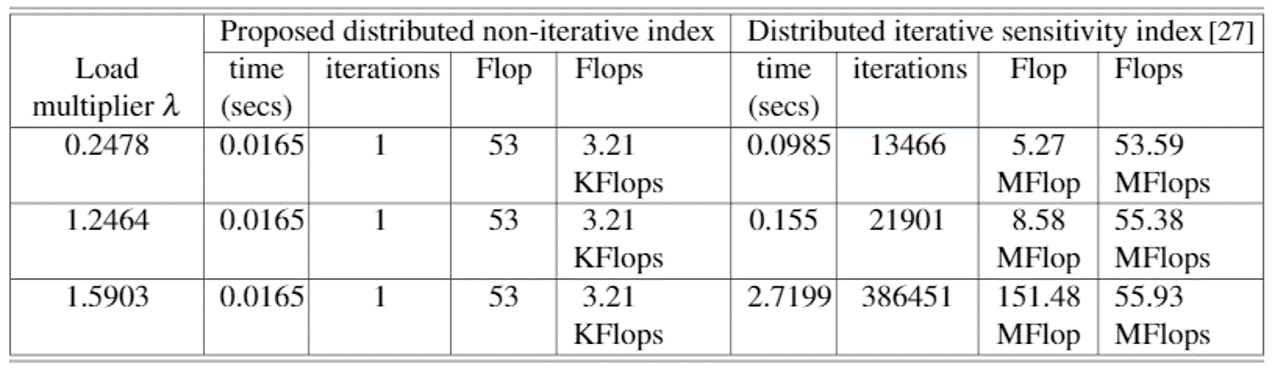}
  \label{tab:flops_comp}
  \vspace{-2em}
\end{table}

\subsection{Behavior of Proposed VSI after Line \& Generator Outage}
\begin{figure}
    \centering
    \centerline{\includegraphics[height=3.5cm, width=\columnwidth]{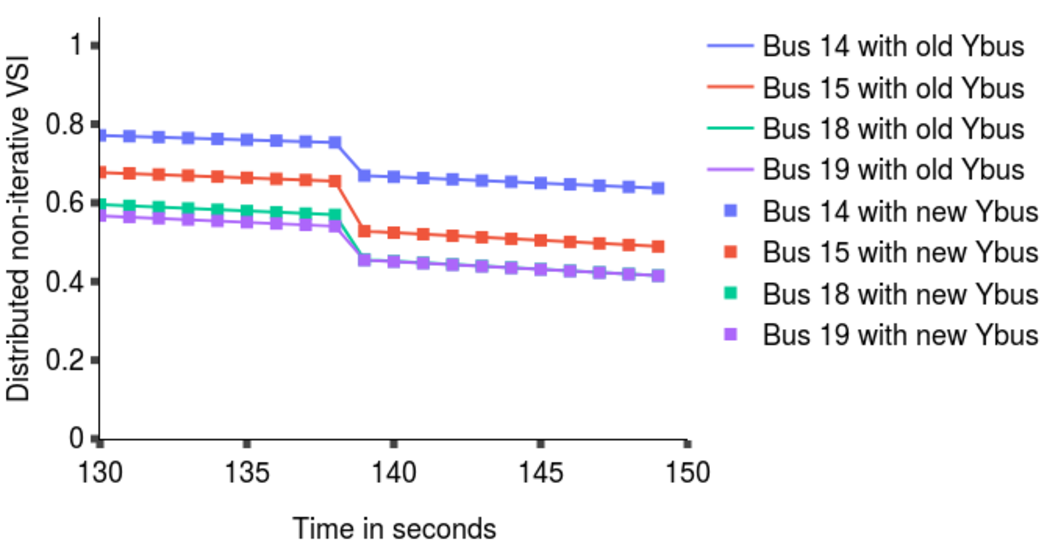}}%Case30_lineoutage.eps
    \caption{Line outage between buses $15$ and $23$ in IEEE-30 bus system. Effect of updating Ybus (new Ybus) and not updating Ybus (old Ybus) on proposed index.}
    \label{fig:line_outage}
    \vspace{-2em}
\end{figure}

Network reconfiguration due to line outage (either due to faults or for maintenance) is a frequent occurrence in the power system. {One key input to the proposed method is the admittance of lines connecting neighboring buses.  In case of a line outage, this information is outdated and so it is important to investigate the behavior of the index in this scenario. We identified that the proposed index accurately tracks the system stress in real-time even when using the outdated line admittances.} In order to study the effect of topology change on the VSI, the line between buses $15$ and $23$ in the IEEE $30$-bus system is taken out of service {at time $t=138$ secs} as the load is increasing. Fig.~\ref{fig:line_outage} plots the VSI at buses $14,15,18$ \& $19$ and it is observed that VSI drops the moment the line outage occurs at time $t=138$ secs, indicating that the {system is stressed and margin to voltage collapse point has reduced. It is also observed from Fig.~\ref{fig:line_outage} that the values of the VSI at various buses are very similar when using un-updated and updated connectivity information. Thus the distributed VSI calculation methodology provides a reasonable estimate of the VSI for the operator/relay to trigger controls/alarms in the time it takes to correct the admittance information, thus improving the situational awareness of the grid.}  In addition, it can also be observed that the index at bus $15$ that is closest to the line outage reduces the most and this fact can be used to identify the fault location in an unsupervised manner.

% \vspace{-1.5em}

\begin{figure}
    \centering
    \centerline{\includegraphics[scale=0.4]{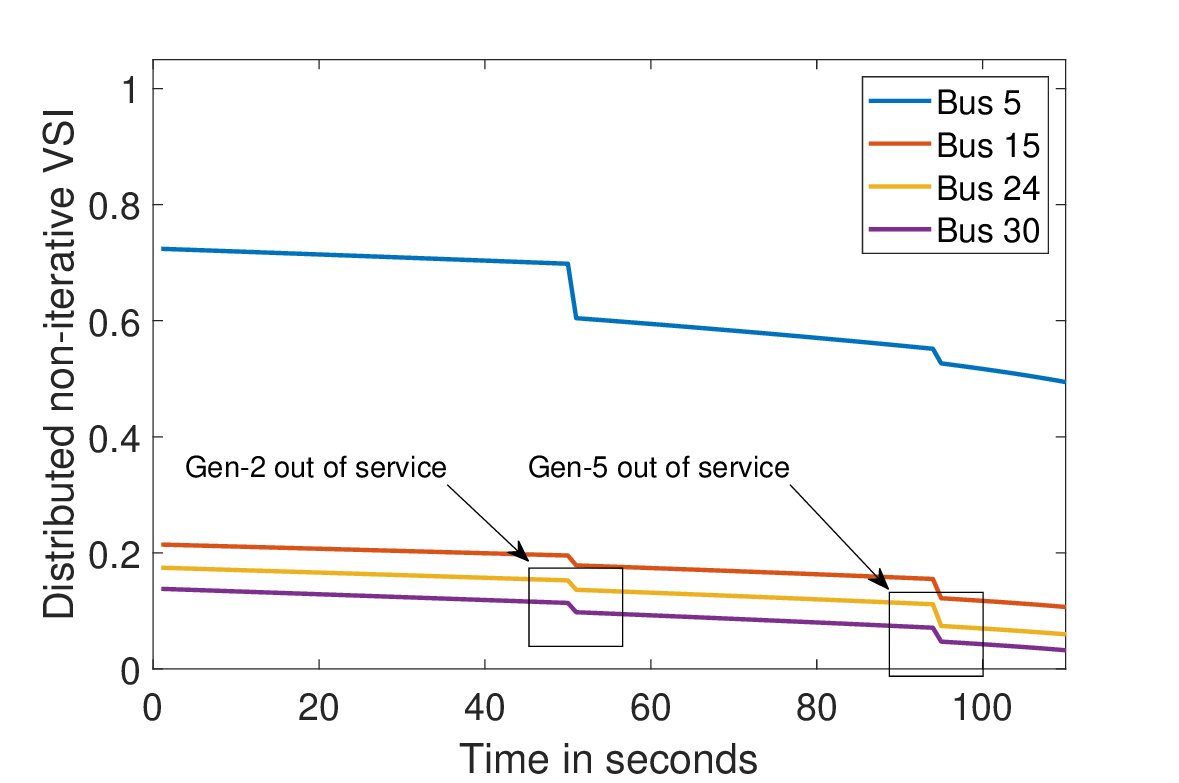}}
    \caption{Behavior of VSI for gen-2 (bus 2) and gen-5 (bus 23) outages in IEEE-30 bus system.}
    \label{fig:gen_outage}
    \vspace{-2em}
\end{figure}
{Another disturbance that is common is the outage of a generator. We investigated this scenario and observed that the proposed VSI accurately tracks the system stress after a generator outage and can identify the critical buses. Consider the IEEE-$30$ bus system, generator two located at bus 2 is outaged at 50 secs and generator five located at bus 23 is outaged at 93 secs. We can observe from Fig.~\ref{fig:gen_outage} that at time t= 50 secs and t=93 secs, VSI at all the buses drop, indicating rising overall system stress. Further, the drop is more at locations closer to the generator with larger power output (gen-2), quantifying the severity of the outage. }

Thus, these results validate the proposed VSI's ability to identify LTVI and demonstrate it's utility to monitor voltage stability using measurements in a distributed manner under various disturbances such as network reconfiguration and generator outage.
	%--
	\section{Conclusion}
	\label{sec:conclusion}
	This paper proposes a PMU measurement-based voltage stability index that can accurately identify long term voltage instability of the system. The key novelty of this work is the mathematical derivation of the index to reflect voltage security and its distributed nature. The index is derived by analyzing the power flow equations in the rectangular form as circles, and at the nose point, the value of the index at the critical bus is zero. In addition, the distributed communication framework between neighboring buses makes the calculation of index scalable and secure for the grid. In the various test scenarios such as noisy measurements, different load increase directions, {line outage and generator outage, the index detected the critical bus and quantified the stress in the system.} The proposed index behavior is also compared with distributed and decentralized methods and is shown to have a superior \& reliable performance, particularly for noisy measurements and large grids. The wide area nature of the proposed index makes it robust to measurement and system noise that adversely affects similar techniques such as local Thevenin methods, etc. Furthermore, the proposed index can localize the line/gen outage locations in an unsupervised manner purely using PMU measurements, making it a promising method for event detection and localization of other events. Finally, the distributed nature of the index makes it possible to utilize cloud computing infrastructure and other recent trends in the big data analytics field for efficient computation and storage. 
	%--
	\bibliographystyle{IEEEtran}
	\bibliography{Kishan}
	%--
	%\onecolumn
	%--
	\appendix
	\label{sec:app}
	\subsection{{Matrix Representation of Standard Circle:}} \label{subsec:app:mat_rep_std_circles}
First, the matrix form of a standard circle is derived. Let $\mathit{z} = \mathit{x}+j\cdot\mathit{y}$ in complex plane be set of all the points in a circle $\mathit{{C_k}\left( z, z^* \right)}$ with radius $\rho$ and center $\gamma = \alpha+j\cdot\beta$. Then, the equation of this circle is given by \eqref{eq:cirle_eq111}
\begin{align}
    &\left(\mathit{x} - \alpha\right)^2 + \left(\mathit{y} - \beta\right)^2 = \rho^2, \label{eq:cirle_eq111}\\
        & \left|\mathit{{z}} - \gamma\right|^2 = \rho^2, \notag \\
    & \left(\mathit{{z}} - \gamma\right)\cdot\left(\mathit{{z}^*} - \gamma^*\right) = \rho^2, \notag\\
    &\mathit{{z}}\cdot\mathit{{z}^*} - \gamma^* \cdot \mathit{{z}} - \gamma\cdot\mathit{{z}^*} + \gamma\cdot\gamma^* - \rho^2 = 0, \label{eq:circle_eq222} \\
    &\triangleq \begin{bmatrix} A & B \\ C & D \end{bmatrix}\ \triangleq  \begin{bmatrix} 1 & -\gamma^* \\ -\gamma & (\gamma\cdot\gamma^* - \rho^2) \end{bmatrix}, \notag
\end{align} 
{where $\mathit{A} = 1$, $\mathit{B} = -\gamma^*$, $\mathit{C} = -\gamma$ and $\mathit{D} = \gamma\cdot\gamma^* - \rho^2$.}

%---------------------------------------------------------------------
\subsection{{Proof for linear behavior of proposed index in case of transmission networks:}}
\vspace{-2.25em}
\label{subsec:app:proof_linearity}
\begin{figure}[H]
    \centering
    \centerline{\includegraphics[scale=0.5]{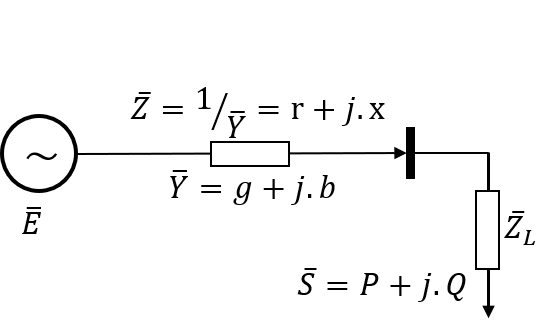}}
    \caption{{$2$-bus system with load connected to an infinite bus.}}
    \label{fig:two_bus_system_rev_4}
\end{figure}

{Fig.~\ref{fig:two_bus_system_rev_4} shows a simple $2$-bus system which consists of one load fed by an infinite bus through a transmission line. The infinite bus is represented by an ideal voltage source with constant voltage given by $\overline{E}=E_r+j \cdot E_i = 1+j\cdot 0$. The transmission line is represented by its admittance $\overline{Y} = g+j\cdot b$. The load consumes an apparent power of $\overline{S}=P+j\cdot Q$ with a constant power factor $cos(\phi)$ and it is represented by a constant equivalent impedance $\overline{Z}_{L} = R_L+j \cdot X_L$.}

{The proposed distributed non-iterative index is given by 
\begin{align}
\Delta^*_{\textrm{norm}} = \dfrac{\Delta^*}{\Delta^*_{\textrm{no-load}}},\label{eq:norm_indexxx_rev_4}
\end{align}
where $\Delta^*$ is given by \eqref{eq:delta_star_rev_4}, $\Delta^*_{\textrm{no-load}}$ is the value of $\Delta^*$ with zero load and voltage of 1 p.u. at all buses of the network.}
{\begin{align}
\Delta^{*} &=  \left(\mathit{c}_p - \dfrac{\left\|\bd b_{p}\right\|^2}{4} \right) \cdot \left(\mathit{c}_q - \dfrac{\left\|\bd b_{q}\right\|^2}{4} \right) - \left(\dfrac{1}{8} \cdot \left\| \bd b_{p} - \bd b_{q} \right \|^2 \right. \notag \\ &\left. - \dfrac{1}{2} \cdot \left(\dfrac{\left\|\bd b_{p}\right\|^2}{4} - \mathit{c}_p \right) - \dfrac{1}{2} \cdot \left(\dfrac{\left\|\bd b_{q}\right\|^2}{4} - \mathit{c}_q \right)\right)^2, \label{eq:delta_star_rev_4}
\end{align}
where
\begin{align*}
    \bd b_{p} &= \begin{bmatrix}\dfrac{t_{d,2}}{t_{d,1}} & \dfrac{t_{d,3}}{t_{d,1}}\end{bmatrix}^{T},\ \bd b_{q} = \begin{bmatrix} \dfrac{-t_{d,3}}{t_{d,4}} &  \dfrac{t_{d,2}}{t_{d,4}}\end{bmatrix}^{T}, \\
    \mathit{c}_p &= \dfrac{-\mathit{p}_{d}}{t_{d,1}},\ \mathit{c}_q = \dfrac{-\mathit{q}_{d}}{t_{d,4}}.
\end{align*}}

{Let load power factor be unity ($cos(\phi)=1$). Now, calculate the terms $\bd b_{p}$, $\bd b_{q}$, $c_p$ and $c_q$ as shown below for the $2$-bus example that is described above.
\begin{align}
    \bd b_{p} &= \begin{bmatrix}-1 & \dfrac{-b}{g}\end{bmatrix}^{T},\ \bd b_{q} = \begin{bmatrix}-1 & \dfrac{g}{b}\end{bmatrix}^{T}, \notag \\
    \mathit{c}_p &= \dfrac{P}{g},\ \mathit{c}_q = 0\ \text{(since unity power factor)}. \notag
\end{align}}

{Substitute above terms ($\bd b_{p},\bd b_{q}$,$c_p,$ and $c_q$) in \eqref{eq:norm_indexxx_rev_4}  and upon further simplification, \eqref{eq:norm_indexxx_rev_4} is given by
\begin{equation}
    \Delta^*_{norm} = \frac{{b}^4+2\,{b}^2\,{g}^2-4\,{b}^2\,g\,P-4\,{b}^2\,P^2+{g}^4-4\,{g}^3\,P}{{\left({b}^2+{g}^2\right)}^2}, \label{eq:simplified_eq_0_rev_4}
\end{equation}}

{\textbf{Bounds of proposed index}: First, we will show that the proposed index is bounded between $1$ and $0$ at no load ($P = 0$) and nose point of the PV curve ($P=P_{max}$) respectively. The maximum transferable power ($P=P_{max}$) for the $2$-bus system from Fig.~\ref{fig:two_bus_system_rev_4} when the load power factor is unity is given by $$P_{max}= \dfrac{ -(b^2 + g^2)g + (b^2 + g^2)^\frac{3}{2}}{(2 b^2)},$$
% $$P_{max}=\dfrac{\|\overline{Z}\|\cdot E^2}{(r + \|Z\|)^2 + (x)^2} = \dfrac{ -(b^2 + g^2)g + (b^2 + g^2)^\frac{3}{2}}{(2 b^2)},$$ 
\begin{align}
    &\textbf{No load condition:}\ \Delta^*_{norm} = \frac{{b}^4+2\,{b}^2\,{g}^2+{g}^4}{{\left({b}^2+{g}^2\right)}^2} = 1, \notag \\
    &\textbf{Full load condition ($P=P_{max}$):}\ \Delta^*_{norm} = 0. \notag
\end{align}}

{\textbf{Linear behavior of proposed index for transmission networks}: Second, we show that even though the proposed index \eqref{eq:simplified_eq_0_rev_4} is a quadratic equation in real power ($P$), the equation behaves linearly for various practical power system operation ranges since the coefficient of ($P^2$) term is negligible.} Specifically, we showcase that the proposed index behaves linearly under various $r/x$ ratios of the transmission line and various load power factors.
\begin{enumerate}[leftmargin=0pt, itemindent=20pt,
labelwidth=15pt, labelsep=5pt, listparindent=0.7cm,align=left]
    \item \textbf{Effect of $x/r$ ratio on the linear behavior of the index:} The $x/r$ ratio is a critical power system parameter that effects the maximum power transfer capability of the grid. For example, a low $x/r$ ratio enables more maximum power transfer when compared to that of a high $r/x$ ratio power line. Fig.~\ref{fig:x_by_r_ratio_pic_rev_4} shows that the proposed index \eqref{eq:simplified_eq_0_rev_4} behaves linearly  for various values of $x/r$ ratios corresponding to different levels of voltage ($22$ kV to $765$ kV) \cite{moon13,c37_standard}. Fig.~\ref{fig:x_by_r_ratio_pic_rev_4} shows that the proposed index behaves linearly for high $x/r$ values (transmission networks) and behaves in a slightly curved fashion for low $x/r$ values (distribution networks). 
    
    For example, when $x/r= 5$ and $x/r=10$ then \eqref{eq:simplified_eq_0_rev_4} can be further simplified to \eqref{eq:simplified_eq_1_rev_4} and \eqref{eq:simplified_eq_2_rev_4} respectively. From \eqref{eq:simplified_eq_1_rev_4} and \eqref{eq:simplified_eq_2_rev_4}, we can see that the high values of $x/r$ makes the coefficient of ``$P^2$" term negligible and hence the linear part of the equation i.e., ``$P$" term dominates resulting in the linear behavior of the proposed index.
    \vspace{-1.25em}
    \begin{figure}
        \centering
        \centerline{\includegraphics[scale=0.255]{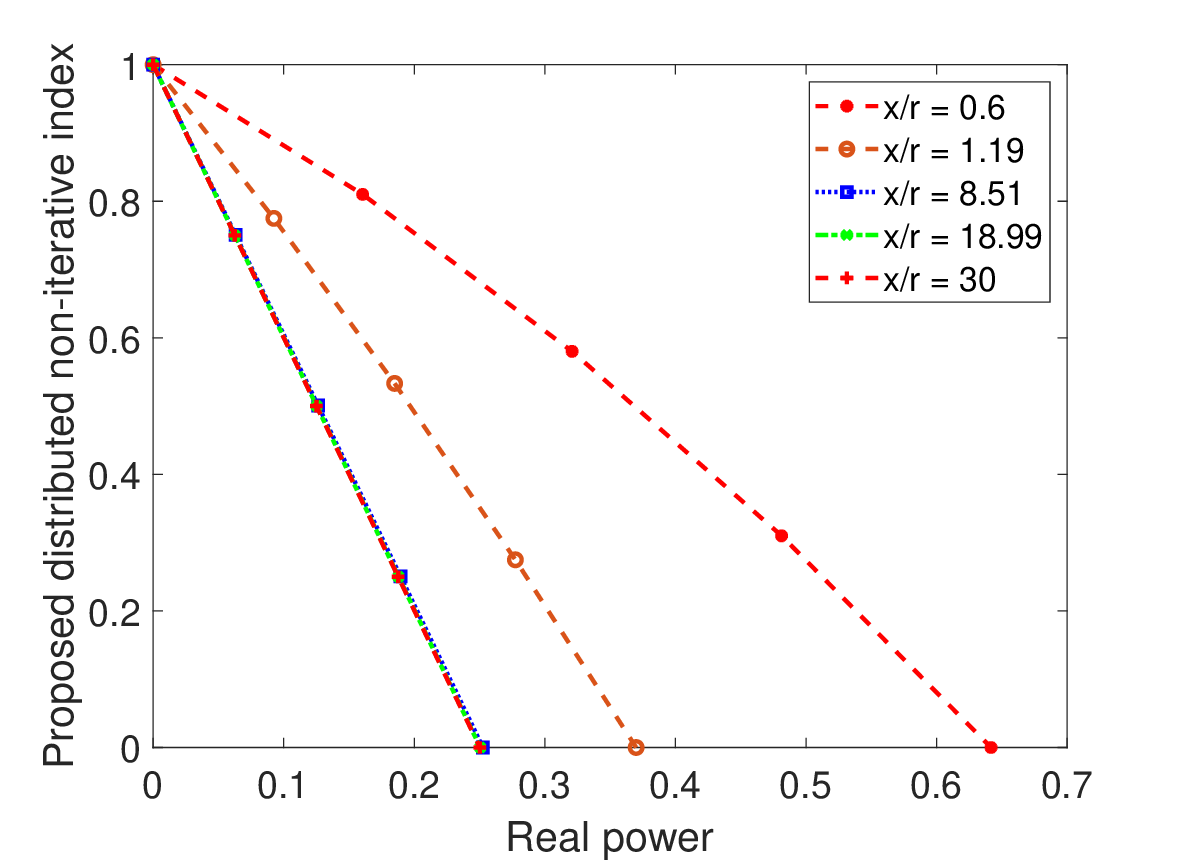}}
        \caption{Proposed index for various values of $x/r$ ratio of the $2$-bus system from Fig.~\ref{fig:two_bus_system_rev_4}. Low $x/r$ ratios such as $0.6,1.19$ represent distribution networks while high $x/r$ ratios such as $8.51,18.99,30$ represent the transmission networks.}
        \label{fig:x_by_r_ratio_pic_rev_4}
    \end{figure}
    \begin{align}
        &\eqref{eq:simplified_eq_0_rev_4}\ \text{(when $x/r=5$)}\ = -\frac{P^2}{169}-\frac{10\,P}{13}+1 \approx -\frac{10\,P}{13}+1. \label{eq:simplified_eq_1_rev_4}
    \end{align}
    \begin{align}
        &\eqref{eq:simplified_eq_0_rev_4}\ \text{(when $x/r=10$)}\ = -\frac{4\,P^2}{10201}-\frac{40\,P}{101}+1 \approx -\frac{40\,P}{101}+1. \label{eq:simplified_eq_2_rev_4}
    \end{align}
    
    Additionally, the proposed index does not deviate a lot from its original slope and hence it tends to stay fairly linear. For example, we know that the first order differentiation of \eqref{eq:simplified_eq_0_rev_4} gives the slope of the proposed index. The second order differentiation of the \eqref{eq:simplified_eq_0_rev_4} gives the information about the change of the slope of the proposed index. From \eqref{eq:simplified_eq_3_rev_4}, when $x/r=10$, this rate of the change of slope of the proposed index is bounded by the coefficient of ``$P^2$" and it is negligible ($-7.8424e^{-4} \approx 0$) for transmission network lines due to high $x/r$ ratio as discussed. In other words, as the $x/r$ ratio increases, the coefficient of $P^2$ becomes negligibly small. Hence the proposed index tends to behave linearly for all the transmission network case studies presented in this paper.
\begin{align}
    \dfrac{d^2(\Delta^*_{norm})}{dP^2}\ = \dfrac{d^2\left(-\frac{4\,P^2}{10201}-\frac{40\,P}{101}+1\right)}{dP^2} = -7.8424e^{-4}. \label{eq:simplified_eq_3_rev_4}
\end{align}
    \item {\textbf{Effect of load power factor on the proposed index}: For a given $x/r$ ratio of $10$, Fig.~\ref{fig:pf_figure_comp_rev_4} shows the proposed index \eqref{eq:simplified_eq_0_rev_4} for various values of load power factor. The proposed index's linear behavior is not effected unless the power factor decreases to as low as $0.6$ which is not realistic for transmission networks.}
        \begin{figure}
        \centering
        \centerline{\includegraphics[scale=0.255]{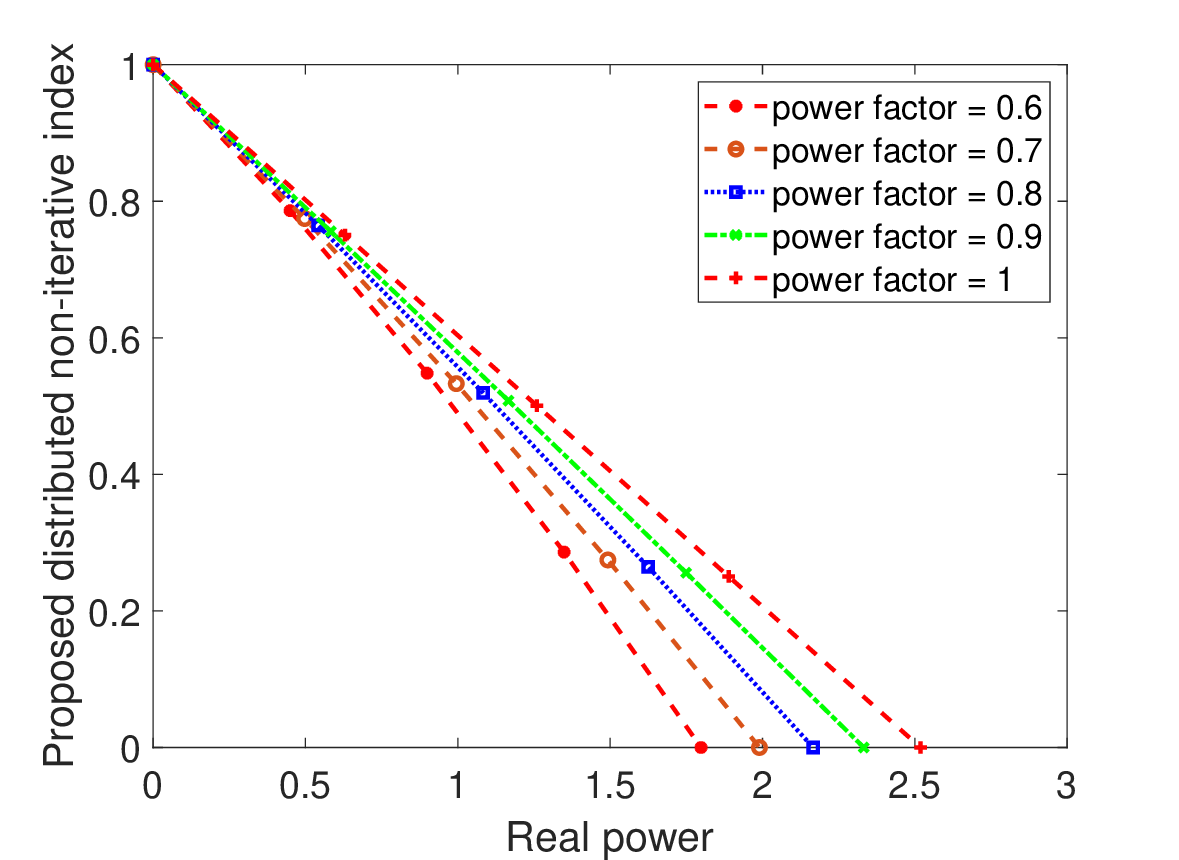}}
        \caption{{Proposed index for various values of the load power factor when the $x/r$ ratio of the transmitting branch is $10$.}}
        \label{fig:pf_figure_comp_rev_4}
    \end{figure}
\end{enumerate}

\end{document}